\documentclass[journal]{IEEEtran}
\usepackage{amsmath,amsfonts}
\usepackage[linesnumbered,ruled,vlined]{algorithm2e}
\usepackage{pifont}
\usepackage{amsthm}
\usepackage{algpseudocode}
\usepackage{tcolorbox}
\usepackage{array}
\usepackage[caption=false,font=normalsize,labelfont=sf,textfont=sf]{subfig}
\usepackage{textcomp}
\usepackage{stfloats}
\usepackage{url}
\usepackage{verbatim}
\usepackage{graphicx}
\usepackage{ulem}
\usepackage{amsfonts}
\usepackage[utf8]{inputenc}
\usepackage{setspace}
\usepackage{cleveref}
\usepackage[utf8]{inputenc}
\crefname{section}{§}{§§}
\Crefname{section}{§}{§§}
\usepackage{caption}
\usepackage{booktabs}
\usepackage{makecell}
\usepackage{MnSymbol}
\usepackage{utfsym}
\usepackage{filecontents, pgffor}
\usepackage{balance}
\usepackage{latexsym}
\usepackage{color}
\usepackage{cite}
\usepackage{listings, xcolor}
\definecolor{verylightgray}{rgb}{.97,.97,.97}

\lstdefinelanguage{Solidity}{
	keywords=[1]{anonymous, assembly, assert, balance, break, call, callcode, case, catch, class, constant, continue, constructor, contract, debugger, default, delegatecall, delete, do, else, emit, event, experimental, export, external, false, finally, for, function, gas, if, implements, import, in, indexed, instanceof, interface, internal, is, length, library, log0, log1, log2, log3, log4, memory, modifier, new, payable, pragma, private, protected, public, pure, push, require, return, returns, revert, selfdestruct, send, solidity, storage, struct, suicide, super, switch, then, this, throw, transfer, true, try, typeof, using, value, view, while, with, addmod, ecrecover, keccak256, mulmod, ripemd160, sha256, sha3}, % generic keywords including crypto operations
	keywordstyle=[1]\color{blue}\bfseries,
	keywords=[2]{address, bool, byte, bytes, bytes1, bytes2, bytes3, bytes4, bytes5, bytes6, bytes7, bytes8, bytes9, bytes10, bytes11, bytes12, bytes13, bytes14, bytes15, bytes16, bytes17, bytes18, bytes19, bytes20, bytes21, bytes22, bytes23, bytes24, bytes25, bytes26, bytes27, bytes28, bytes29, bytes30, bytes31, bytes32, enum, int, int8, int16, int24, int32, int40, int48, int56, int64, int72, int80, int88, int96, int104, int112, int120, int128, int136, int144, int152, int160, int168, int176, int184, int192, int200, int208, int216, int224, int232, int240, int248, int256, mapping, string, uint, uint8, uint16, uint24, uint32, uint40, uint48, uint56, uint64, uint72, uint80, uint88, uint96, uint104, uint112, uint120, uint128, uint136, uint144, uint152, uint160, uint168, uint176, uint184, uint192, uint200, uint208, uint216, uint224, uint232, uint240, uint248, uint256, var, void, ether, finney, szabo, wei, days, hours, minutes, seconds, weeks, years},	% types; money and time units
	keywordstyle=[2]\color{teal}\bfseries,
	keywords=[3]{block, blockhash, coinbase, difficulty, gaslimit, number, timestamp, msg, data, gas, sender, sig, value, now, tx, gasprice, origin},	% environment variables
	keywordstyle=[3]\color{violet}\bfseries,
	identifierstyle=\color{black},
	sensitive=true,
	comment=[l]{//},
	morecomment=[s]{/*}{*/},
	commentstyle=\color{gray}\ttfamily,
	stringstyle=\color{red}\ttfamily,
	morestring=[b]',
	morestring=[b]"
}

\begin{document}
% COBRA: Interaction-Aware Bytecode-Level Vulnerability Detector for Smart Contracts
\title{Interaction-Aware Vulnerability Detection in Smart Contract Bytecodes}

\author{Wenkai Li, Xiaoqi Li, Yingjie Mao, Yuqing Zhang
\IEEEcompsocitemizethanks{\IEEEcompsocthanksitem Wenkai Li, Xiaoqi Li, Yingjie Mao are with the School of Cyberspace Security, Hainan University, Haikou, 570228, China. 
 E-mail: cswkli@hainanu.edu.cn, csxqli@ieee.org, yingjiemao@hainanu.edu.cn;
\IEEEcompsocthanksitem Yuqing Zhang is with National Computer Network Intrusion Protection Center, University of Chinese Academy of Sciences, Beijing, 100049, China.
 E-mail: zhangyq@nipc.org.cn.
}
\thanks{Corresponding author: Xiaoqi Li}
\thanks{This manuscript is an extended version of our work \cite{wenkai2024CobraAse}. It has been extended more than 40\% over the ASE conference version, including: (1) Enhancement of the analysis of the experiments (\cref{sec:experiments}).
(2) Addition of case analysis with the exploits detected by our framework (\cref{sec:experiments}).
(3) Elaboration on the extensive discussion of existing literature (\cref{sec::dis}).
(4) Optimization of the deeper analysis of the detected vulnerabilities (\cref{sec::dis}).} 
}

\markboth{IEEE Transactions on Dependable and Secure Computing, ~Vol.~XX, No.~X, XXXX}%
{Wenkai Li\MakeLowercase{\textit{et al.}}: Interaction-Aware Vulnerability Detection in Smart Contract Bytecodes}


\IEEEtitleabstractindextext{
\begin{abstract}
The detection of vulnerabilities in smart contracts remains a significant challenge. While numerous tools are available for analyzing smart contracts in source code, only about 1.79\% of smart contracts on Ethereum are open-source. For existing tools that target bytecodes, most of them only consider the semantic logic context and disregard function interface information in the bytecodes. In this paper, we propose \textsc{COBRA}, a novel framework that integrates semantic context and function interfaces to detect vulnerabilities in bytecodes of the smart contract. To our best knowledge, \textsc{COBRA} is the first framework that combines these two features. Moreover, to infer the function signatures that are not present in signature databases, we propose \textsc{SRIF}, automatically learn the rules of function signatures from the smart contract bytecodes. The bytecodes associated with the function signatures are collected by constructing a control flow graph (CFG) for the \textsc{SRIF} training. We optimize the semantic context using the operation code in the static single assignment (SSA) format. Finally, we integrate the context and function interface representations in the latent space as the contract feature embedding. The contract features in the hidden space are decoded for vulnerability classifications with a decoder and attention module. Experimental results demonstrate that \textsc{SRIF} can achieve 94.76\% F1-score for function signature inference. Furthermore, when the ground truth ABI exists, \textsc{COBRA} achieves 93.45\% F1-score for vulnerability classification. In the absence of ABI, the inferred function feature fills the encoder, and the system accomplishes an 89.46\% recall rate. 
\end{abstract}

}

\maketitle

\IEEEdisplaynontitleabstractindextext

\IEEEpeerreviewmaketitle

\section{Introduction}
\label{sec:intro}

Detecting vulnerabilities in smart contracts is a crucial task in blockchain systems, which are distributed ledgers that publicly record transactions. Until May 2024, there are about 66 million deployed contracts \cite{2023googlebigQuery}, while only around 1.19 million are open source to the public \cite{2023SmartContractSanctuary}, accounting for approximately 1.79\% of the total. With the advent of the smart contract layer, Ethereum has gained enhanced functionality. However, as the use of smart contracts proliferates, numerous fragile code snippets are exploited maliciously. For example, the reentrancy vulnerability that led to a 3.6M ETH loss in the DAO event \cite{VB2016DAO_Vulnerability}, and the error of arithmetic that caused an \$80M loss in Compound Finance \cite{CompoundFinance2021}.

\begin{figure}[!t]
  \centering
  \includegraphics[width=\linewidth]{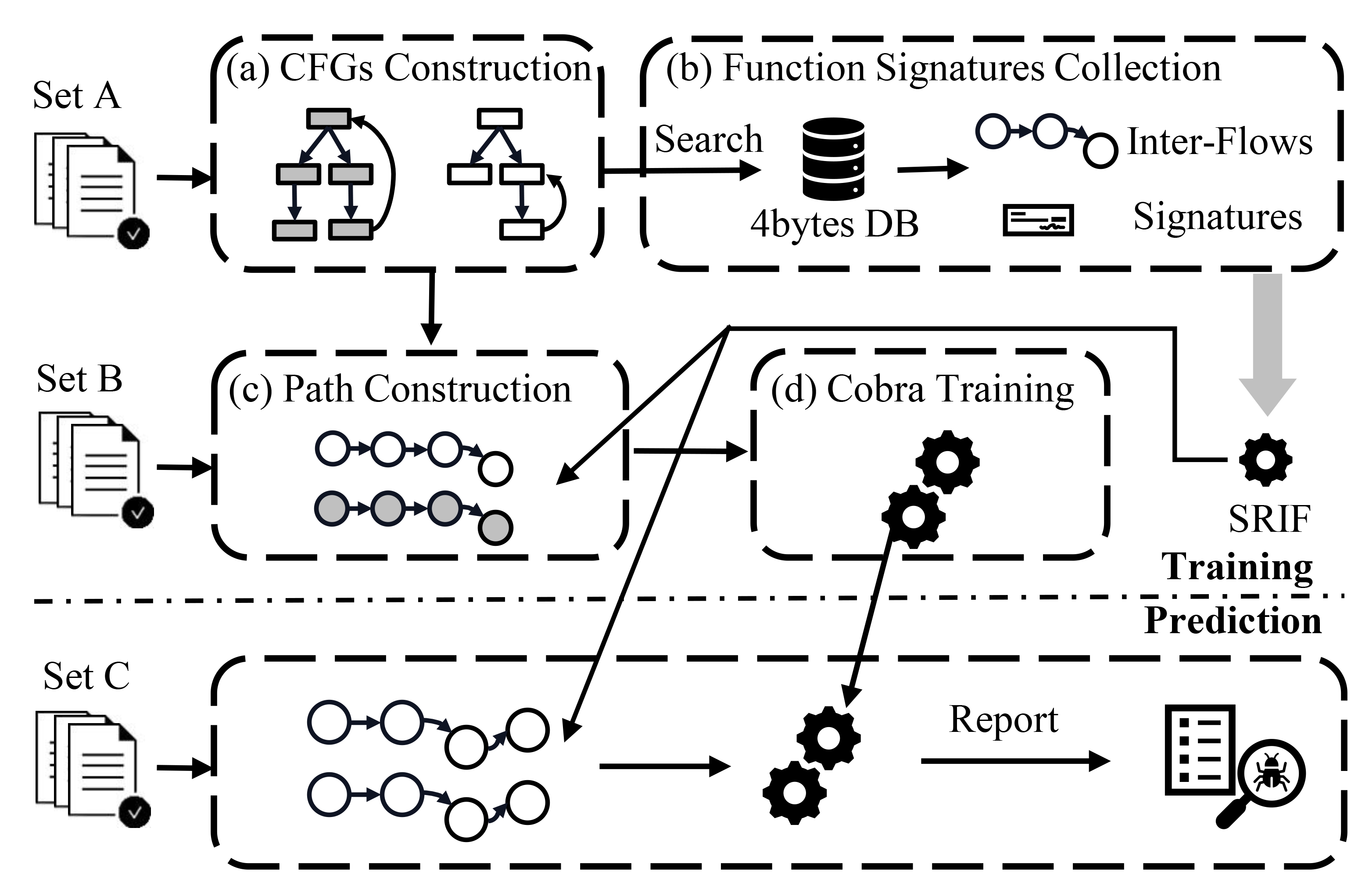}
  \caption{Overview of Framework.}
  \label{fig:overall_framework}
  \vspace{-3ex}
\end{figure}

Previous studies have leveraged several dynamic and static methods to detect contract vulnerabilities. Symbolic execution (e.g., \cite{luu2016making}\cite{mossberg2019manticore}), fuzz testing (e.g., \cite{jiang2018contractfuzzer}\cite{val2020tgFuzz}), and taint analysis (e.g., \cite{rodler2018sereum}) are viable attempts for detecting smart contract vulnerabilities. Moreover, they rely on the control flows within functions to some extent. Symbolic execution leverages formal methods to analyze all variables, including function parameters, to calculate potential sequences of vulnerabilities mathematically \cite{luu2016making}. Fuzzing tests attempt to expose flaws in the contract by using unreasonable inputs, but the presence of function interfaces may diminish their effectiveness \cite{jiang2018contractfuzzer}. Taint analysis tracks and identifies whether tainted source data will be maliciously processed to expose a vulnerability, and functional interfaces can be the source of the taint \cite{rodler2018sereum}. Recent advancements in technology and the availability of large datasets have given rise to new approaches, such as machine learning, which has been used to analyze transactions and accounts to detect Ponzi schemes (e.g., \cite{chen2018detecting}\cite{hu2022scsguard}), and other vulnerabilities (e.g., \cite{gao2020deep}\cite{So2021smarTest}\cite{sendner2023smarter}). 

However, these tools utilized the semantic execution sequence of the source code. They do not prioritize the role of function interfaces in their detection process, even when analyzing bytecode. In addition, the source code of smart contracts can be transformed into bytecode and application binary interface (ABI) after compilation. Therefore, inspired by the deployment and interactions of contracts, we jointly learn the semantic features of contracts and the function interfaces to detect malicious contracts. 

% The symbolic execution technique was one of the earliest attempts to detect vulnerabilities in smart contracts. For example, the \textsc{Oyente} \cite{luu2016making} analyzes the execution traces in the static smart contracts, and bug rules are predefined to confirm the classifications. However, it generates excessive false positives by disregarding specific rules. On the other hand, the \textsc{Manticore} \cite{mossberg2019manticore} validates various parameters by providing customized interfaces, and supports setting the start conditions to increase code coverage. Moreover, invariants in the contract can be identified by tracing through possible state transitions. However, the greater the stack depth, the more resources and time are required. 

% These excellent tools detect vulnerabilities through the semantic sequence at the code level. Recently, several new tools have emerged due to the evolution of technology and data expansion. For example, machine learning has been used to analyze transactions and accounts, identifying Ponzi schemes (e.g., \cite{chen2018detecting}, \cite{hu2022scsguard}), and others (e.g., \cite{gao2020deep}, \cite{So2021smarTest}, \cite{sendner2023smarter}).

In this paper, we propose \textsc{cobra}, as demonstrated in Figure \ref{fig:overall_framework}, a deep learning-based framework that \textit{first} integrates semantic context and function interface for vulnerability detection. As a sequence-to-sequence learning structure, the encoder in \textsc{cobra} consists of the following four key components:

\noindent\textbf{\underline{Part (a):}} A semantic extraction process extracts the semantic context in static single assignment (SSA) format, utilizing control flow graph (CFG) construction from bytecode contracts. \textbf{\underline{Part (b):}} A function signatures recovery component, which contains the application binary interface (ABI), signatures collection, and \textsc{srif}. \textsc{srif} first collects public signatures for training, and then retrieves the undisclosed function interfaces. \textbf{\underline{Part (c):}} A path sequence construction function that concatenates semantics and function interfaces. Function interfaces and properties are connected to form a function embedding.

\noindent\textbf{\underline{Part (d):}} A model training component that learns the vulnerability patterns from the semantic and function representations.

The main contributions of this paper are as follows:
\begin{itemize}
    \item To the best of our knowledge, we are the \textit{first} to propose \textsc{SRIF} utilizing a seq2seq structure to extract function parameters from the semantic context. Moreover, we infer the function properties by counting particular Opcodes, jointly mapping as a function feature (\cref{subsec:sigInference}, \cref{subsec:attrInference}).
    \item As far as we know, we are the \textit{first} to present \textsc{COBRA} that integrates semantic context and function interface features, generating an embedding of smart contracts. The embedding is used to classify vulnerabilities (\cref{subsec:vulnerability_detection}).
    \item We integrate inferred function features and semantic information to discover vulnerabilities. Experimental results show that over 94\% F1-score 
    can be implemented if raw ABI is available, and over 89\% recall can be achieved with the inferred function feature (\cref{sec:experiments}).
    \item We also open source some relevant datasets and codes at \url{https://figshare.com/articles/dataset/22313074}.
    
\end{itemize}

As the extended version of the conference paper \cite{wenkai2024CobraAse}, the following significant extensions are provided. 
\begin{itemize}
    \item We present the first empirical study of function parameter and compiler version distributions across the Ethereum blockchain, while exploring SRIF's detection performance for diverse compiler environments.
    \item To evaluate the utilization degree of computing resources by different models, we select LSTM, Transformer, and BERT to analyze the neuron coverage rate, output accuracy, and model parameters. It is found that LSTM could achieve a 6.95\% point increase in accuracy with at least 8 times fewer parameters than Transformer and BERT.
    \item To explore the opcode distribution in the vulnerable contract bytecode, we leverage COBRA to analyze the opcode frequencies of 5 types of vulnerable contracts. Specifically, we remove the invalid instructions that do not concern data operations using SSA opcodes, enhancing the interpretability of the fragile contract bytecode.
    % Through COBRA-based empirical studies, we identify opcode distribution patterns strongly correlated with five vulnerability types, significantly improving vulnerable contract bytecode interpretability.
    
\end{itemize}

The remainder of the paper is organized as follows. Section \ref{sec:bg} provides the background of the paper and Section \ref{sec:method} details the implementation of \textsc{COBRA}. In Section \ref{sec:experiments}, we show the experimental results to demonstrate the effectiveness of our proposed method. The discussion of \textsc{COBRA} is conducted in Section \ref{sec::dis}. Finally, we review related literature in Section \ref{sec:related_work} and conclude our work in Section \ref{sec:conclusion}.

\section{Motivation \& Background}
\label{sec:bg}

% In this section, we introduce smart contract, Ethereum virtual machine, function interface, and related vulnerabilities.
\subsection{Motivation}
The targeted malicious interaction in this paper is shown in Listing \ref{lst:motivation}, providing an understandable source code format.
In the Solidity contracts, the attacker initially records the victim's address in the \texttt{setAddr()} function (line 12). The attack starts at line 15, with the \texttt{Attacker} depositing funds into the \texttt{Victim} contract and retrieving half via a \texttt{call} operation. Since the \texttt{call} lacks a return function specification, execution proceeds to the fallback function (line 21) without altering the \texttt{balances}. Consequently, line 23 recursively retrieves funds regardless of the check at line 5, resulting in an error when the victim's balance is insufficient. In this process, an attack pattern exploiting vulnerabilities is automated into the \texttt{Attacker} contract. It harnesses the call operation to interact with compatible function interfaces, executing the attack logic. 

\begin{lstlisting}[language=Solidity,
                    float,floatplacement=t,
                    basicstyle=\footnotesize\tt,
                    numbers=left,
                    captionpos=b,
                    aboveskip = 0.5em,
                    belowskip = -1.5em,
                    numbersep = -1em,
                    caption= The Simplified Snippets of a Malicious Interaction,
                    label=lst:motivation,
                    ] 
  contract Victim {
     mapping(address => uint256) balances;
     ...
     function withdraw(address add, uint amount){
        require(balances[add]>amount);
        add.call.value(amount)();
        balances[add] -= amount;
     }
  }
  contract Attacker {
     address victim;
     function setAddr(address add) public{
        victim = add
     }
     function attack() payable{
  //deposit money on Victim with call operation
        deposit_call(money)
        victim.call(bytes4(keccak256("withdraw(address add, uint amount)")), money/2);
     }
     //fallback function
     function () payable{
        //Reentrancy
        victim.call(bytes4(keccak256("withdraw(address add, uint amount)")), this.msg.value);
     }
  }
\end{lstlisting} 

\begin{figure}[ht]
  \centering
  \includegraphics[width=\linewidth]{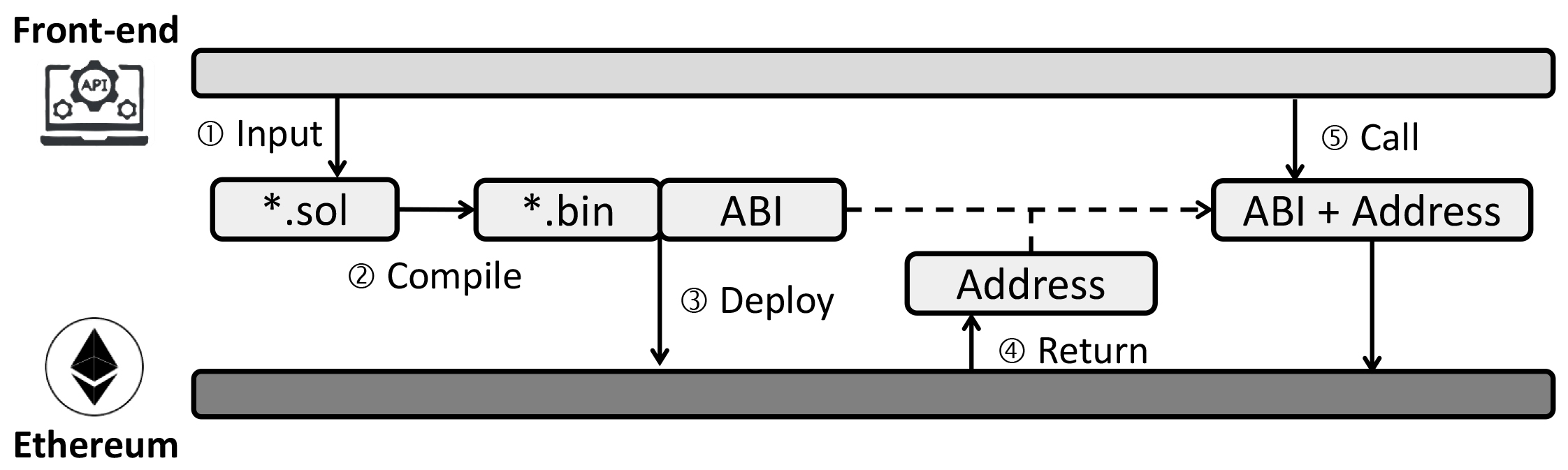}
  \caption{Deployment and Interaction of a Contract. The \ding{172}-\ding{175} represent the whole deployment process of a contract; the \ding{176} and the return of \ding{175} constitute an interactive process.}
  \label{fig:main_idea}
  % \vspace{-2ex}
\end{figure}

As Figure \ref{fig:main_idea} illustrates, the smart contract source code is compiled to produce bytecode and an application binary interface (ABI). The ABI specifies the standardized format for interactions with the contract, including information about functions (e.g., type, name, input parameters, output parameters, and properties). In the Ethereum virtual machine (EVM), a function selector identifies the function signature from various functions based on the first four bytes of the \textsc{calldata}. After the contract is deployed on the blockchain, its address is returned to the front end, which can then use the ABI to call specific contracts.

\subsection{Background}
\subsubsection{Smart Contract}
As a Turing-complete language, smart contract \cite{sailfish2022bose} is initially merged with blockchain in Ethereum. Ethereum's smart contract is compatible with various programming languages, including Solidity, Rust, Vyper, etc. The functional architecture permits smart contracts to communicate with other contracts. The contract is the code deployed on the blockchain, and the deployment process requires only a single transaction containing the compiled code\cite{zapper2022steffen}. Notably, it cannot be modified once the contract code has been released. After deployment, smart contracts can be interacted with each other by invoking with specified function signatures.

\subsubsection{Ethereum Virtual Machine}
The Ethereum virtual machine (EVM) is the execution environment for smart contracts, and nodes in the network can be connected through clients such as Geth \cite{go_ethereum2022}. EVM directly modifies the status information in the state database (StateDB) when a user account initiates a transfer request \cite{chen2022wasai}. If an account submits a transaction request, EVM examines the data field in the message for a function entry to the contract based on the function signature. The interpreter converts the bytecode in StateDB to the Opcode to execute more advanced functionality \cite{ghaleb2022etainter}. The EVM opcodes occupy the hexadecimal bits 0x00-0xFF, with each byte containing only one opcode. These instructions can operate all types of data, including stack data (e.g., \textsc{push}, \textsc{pop}), memory data (e.g., \textsc{mstore}, \textsc{mload}), storage data (e.g., \textsc{sload}, \textsc{sstore}). Further, it can perform arithmetic operations (e.g., \textsc{add}), jump the program counter (e.g., \textsc{jump}), and so on \cite{wood2014ethereum}. Moreover, all operations adhere to the gas mechanism \cite{ghaleb2022etainter}. Each Opcode necessitates a specific quantity of gas to execute. When the required amount of gas exceeds the threshold, the operation will be rolled back \cite{wood2014ethereum}.

\subsubsection{Function Signature} The function signature comprises the function's name and its arguments in the form of \textit{functionName(param1, param2, ...)}. In the event of interactions between contracts, functional signatures become crucial. Since the function name can be defined arbitrarily, the function's behavior depends more on the number of arguments, the type of arguments, and the function id than on its name. The function id can be determined by applying the \textit{Keccak-256} hash algorithm \cite{Keccak256} to the function prototype string \cite{chen2021sigrec} and getting the first 4 bytes. Existing Function Signature libraries, such as the Ethereum Function Signature Database (EFSD) \cite{4byte}, are utilized to extract function ids for their function parameter types and numbers. The function hash is stored in the first four bits of the \textsc{calldata}, and the called contract retrieves which function is called by extracting the function id. With the function hash in the \textsc{calldata}, the EOAs or contract accounts can invoke the bytecode contract through the request from the front end.
\subsubsection{Application Binary Interface} ABI is an interpreter designed to facilitate communication between bytecode smart contracts on EVM \cite{jiang2018contractfuzzer}. Since smart contracts are deployed with bytecode format in Ethereum, ABI decodes bytecode contracts into a human-readable language to facilitate interaction. Each ABI produces the following five components, 1) function types, 2) function names, 3) function input parameters, 4) function output parameters, and 5) function properties. The function types include \textit{constructor}, \textit{fallback}, and \textit{receive}. In Ethereum, the \textit{receive} type identifies a send/receive function, indicating that the function can receive and transfer Ether. A contract may contain only one \textit{fallback} function with no parameters or return values. The \textit{fallback} function is executed when the call request is not sent to any function of a contract. When a contract is created, its \textit{constructor} function is called to initialize its state. 
% After \ding{175} returns the contract address in Figure \ref{fig:main_idea}, and then diverts the function hashes and parameters into \textsc{calldata}. Finally, the \textsc{calldata} would be sent through a transaction.

\subsection{Related Vulnerabilities}
When smart contracts expand the programmability of blockchain systems, security problems also increase. The Decentralized Application Security Project (DASP) \cite{DASP01} is a project classifying smart contract vulnerabilities based on actual impact. In this paper, we will concentrate on five of these vulnerabilities in Table \ref{tab:related_vulnerabilities}. 

\begin{table}[ht]
\small
  \caption{Related Vulnerabilities in \textsc{DASP}}
  \label{tab:related_vulnerabilities}
  \begin{tabular}{l l}
    \toprule
    \textbf{Categories} & \textbf{Alias} \\
    \midrule
    Reentrancy Vulnerability   & Recursive Call\\
    Arithmetic Vulnerability  & Overflow, Underflow\\
    Unchecked Low Level Calls & Unchecked Send\\
    Transaction Ordering Dependency & Front-Running, TOCTOU \\
    Time Manipulation & Timestamp Dependency\\
  \bottomrule
  % \vspace{-4ex}
\end{tabular}
\end{table}

A variety of works have been yielded for studying these attacks in Table \ref{tab:related_vulnerabilities} \cite{luu2016making, he2019ILFuzzer}. The reentrancy can be discovered \cite{VB2016DAO_Vulnerability} when multiple recursive calls are made to withdraw assets before updating the balance state. The integer overflow, floating-point precision loss, and division by zero are all arithmetic vulnerabilities. Developers risk compromising the program's security if they fail to verify the variables' scope. Unchecked low-level calls occur when the return values of the calls are not effectively handled in the contract, resulting in coin loss \cite{chen2020defing_defets}. The Transaction Ordering Dependency (\textsc{tod}) is also known as Time-Of-Check vs Time-Of-Use (\textsc{toctou}) \cite{2023txtIvanov}. By giving higher gas, the miners were incentivized to preempt other transactions, resulting in alterations to the initial states of the contract. A time manipulation vulnerability exists when a timestamp within a block is exploited to trigger a security event. The smart contract has access to the variables in block (e.g., \texttt{timestamp}, \texttt{difficulty}), the \texttt{block.timestamp} can be modified to cause unexpected issues when many contracts call it simultaneously \cite{perez2021smart}.

\begin{figure*}[ht]
  \centering
  \includegraphics[width=\linewidth]{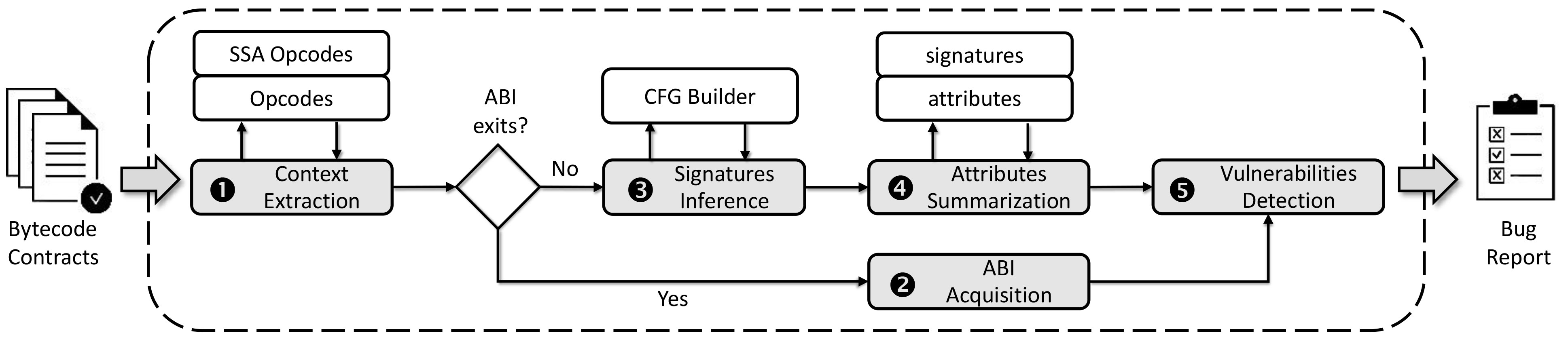}
  \caption{The Architecture of Our Work. The dashed box represents the primary steps of the detection process. The inputs are the bytecode smart contracts, and the output is a vulnerability report.}
  \label{fig:model_overview}
\end{figure*}

\subsection{Neuron Coverage}
Neuron Coverage (NC) \cite{Deepxplore2017NC} quantifies the proportion of activated neurons in a neural network when processing a given test suite. Formally, it is defined as \cref{eqa:NCoverage},

\begin{equation}
NC = \frac{|{ n \in N | \forall i \in T, A(n,i) > \theta }|}{|N|}
\label{eqa:NCoverage}
\end{equation}
where $N$ denotes the complete set of neurons in the network, $T$ represents the collection of test inputs, $A(n,i)$ indicates the activation value of neuron $n$ when processing input $i$, $\theta$ is the predefined activation threshold. A neuron is considered activated when its output value exceeds the specified threshold $\theta$ for any given test input. The metric ranges from 0 to 1, with higher values for a more complete neuron-level test coverage.

\section{COBRA}
\label{sec:method}

In this section, we describe the primary methods taken to implement our two-stage approach, as well as the vulnerability detection model in Figure~\ref{fig:model_overview}. As the input to our approach, the bytecode smart contract performs the following steps: \ding{182} \textit{Context Extraction}, \ding{183} \textit{ABI Acquisition}, \ding{184} \textit{Signatures Inference}, \ding{185} \textit{Attributes Summarization}, and \ding{186} \textit{Vulnerabilities Detection}. 

During operations in \ding{182}-\ding{183} or \ding{182}-\ding{184}-\ding{185}, the dataset for detection model is processed. 
In step 1, the context information (i.e., Opcodes and SSA Opcodes) is extracted from the bytecode smart contracts. In the meantime, we crawl Etherscan \cite{etherscan2023io} for its raw ABI data based on the address of the contract. If the ABI information is collected, both the semantic information and the ABI are represented as embedding, feeding our encoder module to obtain the contract's hidden representation. If the ABI data is absent, the steps \ding{184} and \ding{185} aim to infer the function signatures and attributes to recover the function features in ABI based on semantical context. The function signature includes the function name and parameters. More specifically, since there are only finite function signatures in EFSD, also known as \textsc{4byte}, we propose the \textsc{SRIF} structure for inferring function signatures of contracts. The \ding{186} is the vulnerabilities detection module, where the \textsc{COBRA} with a novel encoder was presented to combine the processed semantic context and function representations of contracts. Finally, a bug report is generated, which contains classifications of the vulnerabilities in the bytecode smart contracts.

\normalem
\begin{algorithm}[ht]
\footnotesize
        \caption{Functions Context and Ids Acquisition}\label{algo:getFunctionOpcde}
	\SetKwData{pushs}{pushs}
        \SetKwData{Block}{block} \SetKwData{Blockins}{block.ins} 
        \SetKwData{opSeq}{OpSeq} 
        \SetKwData{opSeqdest}{OpSeq[dest]} \SetKwData{opSeqeb}{OpSeq[eb]} 
        \SetKwData{opSeqaddr}{OpSeq[fnAddr]} \SetKwData{opSeqentry}{OpSeq[eb]}
        \SetKwData{BasicBlockseb}{BasicBlocks[eb]} 
        \SetKwData{BasicBlocksdest}{BasicBlocks[dest]}
        \SetKwData{BasicBlocksFnAddr}{BasicBlocks[fnAddr]}
        \SetKwData{fnAddr}{fnAddr} \SetKwData{fnid}{fnId}
        \SetKwData{BasicBlocks}{BasicBlocks} \SetKwData{previousPushValue}{prePushValue} \SetKwData{pushValue}{pushValue}  \SetKwData{entryBlock}{eb}
        \SetKwData{IDaddr}{Ids[fnAddr]}  \SetKwData{IDdest}{Ids[dest]}  
        \SetKwData{Dest}{dest}  \SetKwData{IDeb}{Ids[eb]} 
        \SetKwFunction{getFunctionsInfo}{getFuncInfo} 
	\SetKwInOut{Input}{input}\SetKwInOut{Output}{output}
        \SetKwProg{Fn}{Function}{:}{}
        % \DontPrintSemicolon
	\Input{A deployed bytecode smart contract $bc$}
	\Output{two global map: functions context $OpSeq$, functions hashes $Ids$}
	\BlankLine
 
        \emph{\BasicBlocks, \entryBlock$\leftarrow$ CFG.countBasicBlocks(EVMAsm($bc$))};\par
        \emph{\pushValue, \previousPushValue $\leftarrow None$};
        
        \Fn{\getFunctionsInfo{\Block, $entry$}}{ 
            \lForEach{instruction $i$ of the \Blockins}{$Ops$ $\leftarrow$ $i$}
            \If {$entry$}{
            \uIf{end of $block$ compatible with $JUMPI$}{
                 $Assert$ length of \Blockins $>$ 2;\par  
                 \Dest $\leftarrow$ oprand of block.ins[-2];\par 
                 \opSeqdest $\leftarrow$\getFunctionsInfo{\BasicBlocksdest, $false$};\par
                 \IDdest $\leftarrow None$;
             }
            \KwRet $Ops$ \;
            }
            \For{$i$ in \Blockins}{
                \If{$i$ compatible with PUSHs}{ 
                    \previousPushValue $\leftarrow$ \pushValue;\par
                    \pushValue $\leftarrow $ oprand of $i$;
                }
            }
            \eIf{end of $block$ compatible with $JUMPI$}{ 
                \If{\previousPushValue}{
                    \fnAddr, \fnid $\leftarrow$ \pushValue, \previousPushValue;
                }
            }{
                \fnAddr, \fnid $\leftarrow None$;
            }
            \If{\fnAddr compatible with \BasicBlocks}{
                \opSeqaddr $\leftarrow$ \getFunctionsInfo{\BasicBlocksFnAddr, $false$};\par
                \IDaddr $ \leftarrow$ \fnid;\par
                \If {end of \Block compatible with $JUMPI$}{
                    \Dest $\leftarrow$ ((endPc $ep$ of \Block) + 1);\par 
                    \opSeqdest $\leftarrow$ \getFunctionsInfo{\BasicBlocksdest, $false$};
                }
            }
            \KwRet $Ops$ \;
        }
        \ForEach{entryBlock address \entryBlock of the \BasicBlocks}{ 
            \opSeqeb$\leftarrow$\getFunctionsInfo{\BasicBlockseb, $True$};
        }
\end{algorithm}
\vspace{-2ex}

\subsection{Context Extraction}
\label{subsec:contextExtraction}
We first extracted the Opcode in both original and SSA format from the bytecode smart contracts. By decompiling the bytecode with reverse engineering method \cite{crytic2020pyevmasm}, the Opcodes of contracts without compilation error can be gathered. SSA Opcode is an intermediate representation of Opcode that preserves the semantic information by removing data operations in the stack such as \textsc{push}, \textsc{pop}, \textsc{swap}, and \textsc{dup} \cite{crytic2020rattle}.
Then, we construct the CFG in the execution order to obtain Opcodes. It entails separating the contract into basic blocks by searching for instructions about the end of basic blocks. For example, the Opcodes around jump (e.g., \textsc{jump}, \textsc{jumpi}), and others (e.g., \textsc{stop}, \textsc{selfdestruct}, \textsc{return}, \textsc{revert}, \textsc{invalid}, \textsc{suicide}) will result in the stop of sequential execution. 
The resultant basic blocks will execute sequentially, beginning with the first instruction as the entry point and concluding with the last instruction as the outlet.
The CFG is completed by constructing edges between the blocks based on the control flow.
There are 3 types of basic block conversion, including conditional jump (e.g., \textsc{jumpi}), unconditional jump (e.g., \textsc{jump}), and fall to next block. As a function block, the execution block must be processed from its entry. In addition, the function selector requires the block ending with \textsc{jumpi} to obtain the function hashes, identifying different functions.

In the Algorithm \ref{algo:getFunctionOpcde}, the presence or absence of the final instruction \textsc{jumpi} is used to determine whether to proceed to the following function in the program. With the CFG construction, the function hashes and Opcodes of each sequentially executed function block are obtained. For each block, we evaluate its eligibility as an entry for a function. Specifically, on line 5, if the block is an entry and the end instruction is \textsc{jumpi}, we extract the next function id. The Opcodes of the next function are then stored at $OpSeq$, and the address is the key. At line 12, if not an entry, we retrieve the last two push values \texttt{pushValue}, \texttt{prePushValue} within the block. On line 17, if the block ends with a \textsc{jumpi}, the \texttt{pushValue} and \texttt{prePushValue} are used as the address and its hash value. 
% If the address is in the blocks, the Opcodes and the hash in the block will be stored.

Furthermore, the precision of the EVM CFG builder \cite{2023Crytic_EVMCFG} is crucial to the accuracy of the data we collected. For this reason, we keep an optimized EVM CFG recovery module in Elysium \cite{ferreira2022elysium} to gather data for model training. The method in Elysium is proposed by \cite{contro2021ethersolve} \cite{BENINI2023111653}, which would extract more precise CFGs in this module.

\begin{figure*}
    \centering
    \includegraphics[width=1\linewidth]{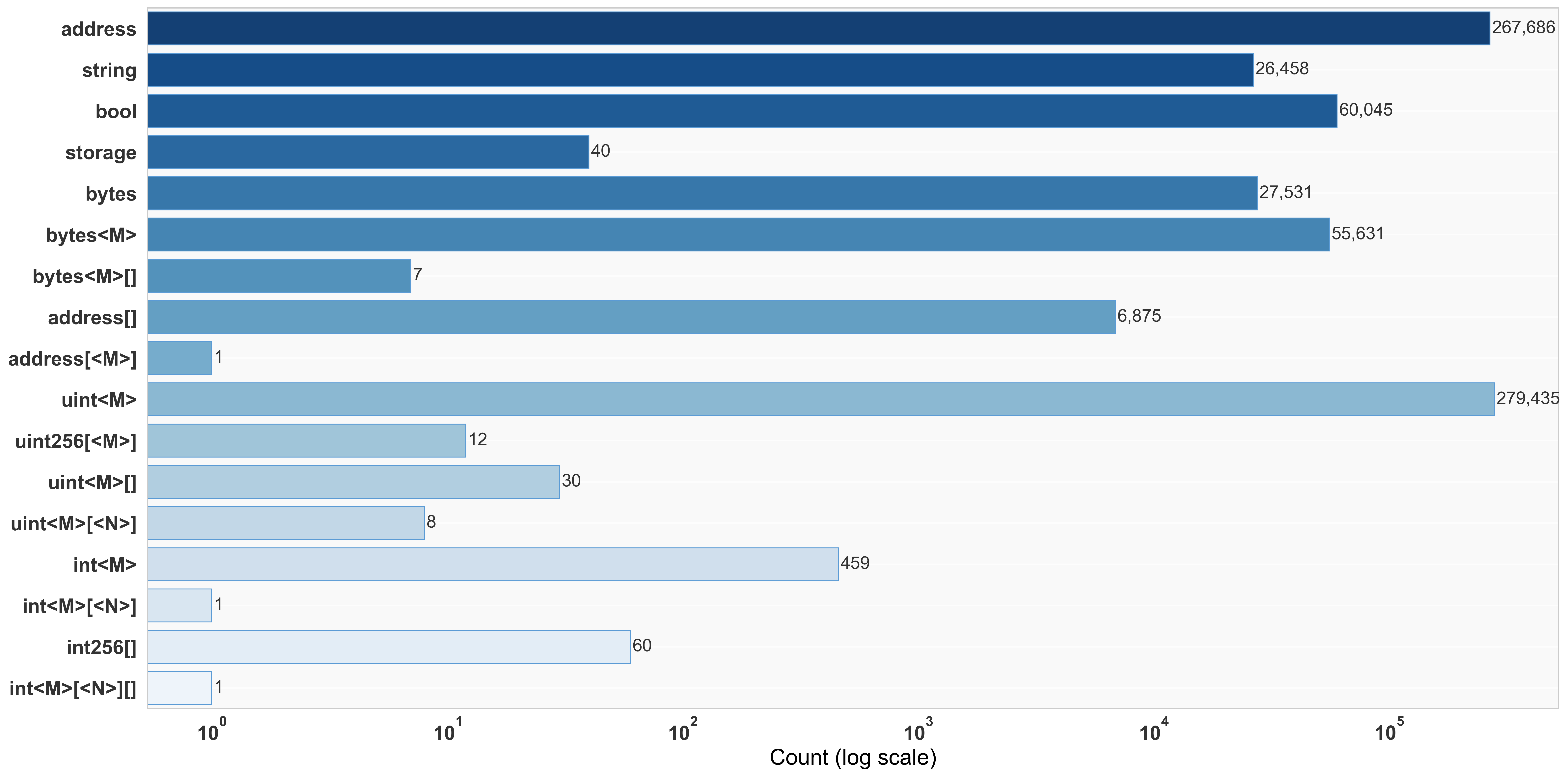}
    \caption{The Label Distribution on the Ethereum Blockchain. The $<$M$>$ and $<$N$>$ are placeholders for a particular data length. The x-scale represents the number of occurrences of the type, which is measured in log to prevent long-tailed effects in the data plot. The y-scale is represented by all parameter types included in the statistics. }
    \label{fig:label_type_distribution}
\end{figure*}

\subsection{ABI Acquisition}
\label{subsec:abiAcquisition}
Etherscan, an Ethereum blockchain browser, allows us to crawl ABI information via contract addresses to examine real-time information such as blocks, transactions, miners, accounts, etc. Similarly to the accessibility of contracts, some bytecode contracts do not make their ABI information publicly available. To demonstrate this, we gathered 96,200 bytecode contracts in block-number order, of which 15,026 ABIs are available. Only 15.62\% of these contracts have published ABI information. Therefore, for these bytecode contracts that do not disclose their ABI information, we propose an alternative approach in \cref{subsec:sigInference} and \cref{subsec:attrInference} for obtaining the function inputs and function attributes from the contracts. The format of the ABI stored in the blockchain is \textsc{json}. To ensure compatibility with the following format for function parameters and properties, we remove the function names.
% convert it to \textsc{tensor} format and

Considering the label space of the function parameters, we collect the function interfaces in the first two million blocks on the Ethereum blockchain, and the number distribution of each parameter type is summarized. As Figure \ref{fig:label_type_distribution} shows, we only collected 17 types of parameters in total, and the address, string, bool, bytes, and uint types appeared more frequently.

\subsection{Signature Inference}
\label{subsec:sigInference}
This section focuses primarily on deducing function parameters information from the function context obtained in the \cref{subsec:contextExtraction}. Although \textsc{4byte} contains publicly available function signatures, the library is incompatible with non-public function ids. In order to initially infer function parameters from bytecode information, we employ a structure called \textsc{SRIF}. 
% Future work would then optimize the vulnerability detection model as an end-to-end framework, motivating our utilization of deep learning technology. However, we just infer these parameters tentatively in this paper.

\begin{figure*}[ht]
  \centering
  \includegraphics[width=\linewidth]{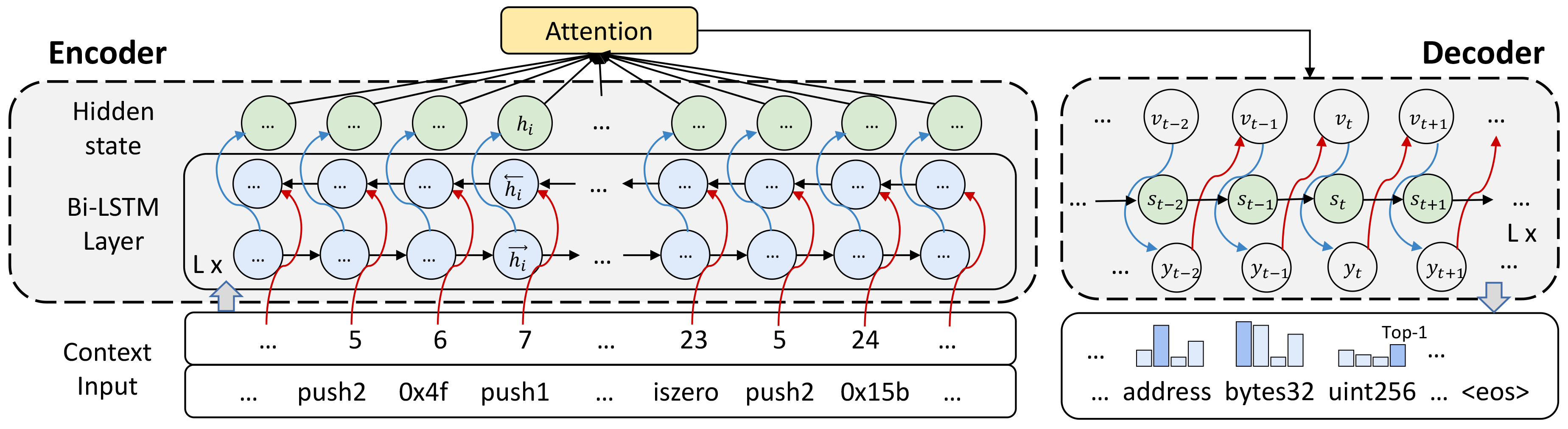}
  \caption{The Signatures Inference Model, \textsc{SRIF}. The dashed boxes represent the encoder and decoder, which are connected by an attention module.}
  \label{fig:model_signature}
\end{figure*}

In accordance with the architecture of the EVM, this stack virtual machine does not store the runtime data, and the function and data operations are embedded within the Opcode. As described in \cref{subsec:contextExtraction}, the function id would be saved in the call data and transferred into the called function. Some Opcodes can interact with call data, e.g., \textsc{calldataload}, \textsc{calldatasize}, and \textsc{calldatacopy}. Simultaneously, these data contain fixed rules for function parameter encapsulation, which can be inferred using specific rules. For instance, over 30 rules were included in \textsc{SigRec} \cite{chen2021sigrec} for deducing function signatures from the bytecode of Solidity and Vyper. Such type-specific inferring rules may be affected by Solidity's version iterations for variable types, and Solidity has experienced more than 100+ released version updates over the past few years \cite{2023Alex-solidityreleases}. Therefore, to achieve a more adaptable process to get these input types in smart contracts, we utilize an encoder-decoder framework in Figure \ref{fig:model_signature} to infer function parameters from the function context.

In \cref{subsec:contextExtraction}, we have gathered the Opcodes of basic block and function id for each function. Subsequently, we employ CFGBuilder to establish the control flow graph (CFG) among all basic blocks. A depth-first search (DFS) algorithm with a designated depth was utilized to capture the sequence of contexts for all basic blocks within the function. We approached the task of parameter prediction as a multi-label classification (MLC) problem, considering the abundance of parameters and label categories involved. As shown in Figure \ref{fig:label_type_distribution}, because the number of smart contract function parameter types is not high, the label space explosion problem of MLC does not occur.

Given the labels $L = \{l_1, l_2, ..., lm\}$, the primary objective is to generate the optimal sequence of label subset $y^*$ from each sentence $\{w_1, w_2, ..., w_n\}$. A subset $s$ of $m$ labels is constructed from $L$ to $x$. The task can be defined as maximizing $P(y|x)$, which can be computed by the following \cref{eqa:seqGen}.

\begin{equation}
    P(y|x)  = \prod_{i=1}^n p(y_i|y_1, y_2, ..., y_i, x)
\label{eqa:seqGen}
\end{equation}

In Figure \ref{fig:model_signature}, we take a context with $n$ words $w_1, w_2, ..., w_n$ as example.  
As input, the word format must be converted to a machine-readable format. We collected possible words as vocabulary $|\nu|$. Suppose $i \in [0, n]$, where $w_i$ is converted to a number, and then one-hot encoding is applied to $w_i$. An embedding matrix $E \in \mathbb{R}^{k \times |\nu|}$ extends each encoded $w_i$ into a $k$-dimensional embedding vector $e_i$. Then the input context can be expressed as $c = \{e_1, e_2, ..., e_n\}$.

To extract the semantic information of each word bidirectionally, we adopt a recurrent neural network, LSTM \cite{hochreiter1997long}, to acquire the context word meaning and its semantic features. The hidden state $h_i$ of each word embedding vector $e_i$ can be obtained by computing \cref{eqa:BiLSTM}, where $h_i$ is the concatenation of the hidden states from both directions, indicating the ultimate representation of the $i$-th word.

\begin{equation}
\begin{aligned}
    % \overrightarrow{h_i} &= \overrightarrow{LSTM}(\overrightarrow{h_{i-1}},c_i),\\
    % \overleftarrow{h_i} &= \overleftarrow{LSTM}(\overleftarrow{h_{i-1}},c_i),\\
    {h_i} = [\overrightarrow{LSTM}(\overrightarrow{h_{i-1}},c_i);  \overleftarrow{LSTM}(\overleftarrow{h_{i-1}},c_i)]
\label{eqa:BiLSTM}
\end{aligned}
\end{equation}

When invoking a function method, the number and sequence of parameters must be correct. Therefore, it is necessary to extract the correct number and order of labels from the original sentence features. In addition, the attention mechanism can identify valid words advantageous to the result from the input sentence. To predict the classification from the hidden states to the variable length, we employ a decoder module with an attention mechanism. Specifically, the context feature vector $v_t$ obtained from attention at time step $t$ is calculated as the \cref{eqa:attention1}, \cref{eqa:attention2}. Where $w_{ti}$ is the weight of $i$-th word at $t$. $\boldsymbol{W_a^T}$, $\boldsymbol{U_a}$, $\boldsymbol{O_a}$ are determined during the training phase.

\begin{equation}
\begin{aligned}
    w_{ti} = softmax(\boldsymbol{W_a^T}\tanh(\boldsymbol{U_a}{s_t} + \boldsymbol{O_a}h_i))
\label{eqa:attention1}
\end{aligned}
\end{equation}
\vspace{-2ex}
\begin{equation}
\begin{aligned}
    v_t = \sum_{i=1}^n w_{ti} h_i
\label{eqa:attention2}
\end{aligned}
\end{equation}
where $s_t$ is the hidden state of decoder at $t$, which can be defined as following \cref{eqa:decoder},

\begin{equation}
\begin{aligned}
    s_t = LSTM(s_{t-1}, [g(y_{t-1});v_{t-1}])
\label{eqa:decoder}
\end{aligned}
\end{equation}
where the $g(y_{t-1})$ represents the label with the highest probability of distribution $y_{t-1}$ at previous time step $t-1$. Notably, the probability distribution $y_{t}$ is generated by a liner layer and a $softmax$ function. Specifically, the hidden states are transformed to the output size using a linear layer with an activation function. The resulting output is then passed through a $softmax$ function to obtain the probability distribution $y_{t}$.

In the training phase, we employ the focal loss function \cite{lin2017focal}. In our multi-classification task, an imbalanced category distribution may hinder the training process and prevent the model from converging to the extremes. The focal loss function was initially introduced for object detection tasks in the computer vision domain, where identifying positive and negative samples may present a wide disparity of difficulty. The focal loss is calculated as \cref{eqa:focalLoss}.

\begin{equation}
\begin{aligned}
    loss = -\alpha(1-p_t)^\gamma log(p_t)
\label{eqa:focalLoss}
\end{aligned}
\end{equation}
where the $\alpha$ controls the weight of positive and negative samples on loss, while $p_t$ represents the probability of the ground truth category. $\gamma$ is also a parameter that controls the value of $(1-p_t)$ to reduce the model's emphasis on easy-to-classify samples close to the ground truth. We followed the original assumption \cite{lin2017focal} that $\gamma$ = 2 for focal loss. For the value of $\alpha$, we made a few minor adjustments so that each class has a more balanced concentration. In \cref{eqa:focalLoss_fineT}, the $n_i$ is the number of types for the $i$-th parameter, $0<i \leq T$, where $T$ is the total number of classes. The $\alpha_i$ value for class $i$ is $ \sum_{j=0}^T n_j / {n_i}$. Each $\alpha_i$ is substituted for $\alpha$ to determine the loss value of each class, and the average value is the total loss. 
% $\frac {\sum_{i=0}^T n_i } {n_i}$
\begin{equation}
\begin{aligned}
    loss' = \frac {\sum_{i=0}^n -\alpha{_i}(1-p_t)^\gamma log(p_t)} {n}
\label{eqa:focalLoss_fineT}
\end{aligned}
\end{equation}
\vspace{-2ex}
\begin{equation}
\begin{aligned}
    \alpha_i = \frac {\sum_{j=0}^T n_j} {n_i}
\label{eqa:focalLoss_fineT1}
\end{aligned}
\end{equation}

\subsection{Attributes Summarization}
\label{subsec:attrInference}
Additionally, we deduce the attributes (i.e., the state mutability and payable) of functions in Solidity. State mutability indicates whether the states of a function can be updated, i.e., functions that are neither \textit{pure} nor \textit{view} types. \textit{View} functions imply that the function state cannot be modified. Prior to the Solidity 0.5.0 version, \textit{view} functions were referred to as \textit{constant}. The \textit{payable} property must be declared when a function transfers or receives Ether in Ethereum. To differentiate these properties and apply them to vulnerability detection, we classify them as \textit{constant}, \textit{pure}, and \textit{payable}. According to Table \ref{tab:statesOpcde}, we summarize the Opcodes that can modify state variables and transfer Ether. For each function, the context is examined for state modification operations and call messages with Ether. Consequently, the \textit{constant} and \textit{payable} properties are inferred, respectively. In particular, functions declared as \textit{pure} neither modify nor read any state variables and, therefore, consume no gas. Based on Ethereum gas consumption\cite{wood2014ethereum}, we get the Opcodes related to the \textit{pure} property of functions.

\begin{table}[ht]
\small
  \caption{The Table of States Operation Opcode}
  \label{tab:statesOpcde}
  \begin{tabular}{p{1.5cm} p{2.7cm} p{2.7cm}}
    \toprule
    \textbf{\makecell[l]{Checked\\Attributes}}& \textbf{\makecell[l]{Related\\Behaviors}} & \textbf{\makecell[l]{Analyzed\\Opcodes}} \\
    \midrule
    View & Storage Modification & \textsc{sstore}  \\
    & Event Emission & \textsc{log0}, \textsc{log1}, \textsc{log2}, \textsc{log3}, \textsc{log4}\\
    & Child Contract Creation & \textsc{create}, \textsc{create2} \\
    & Self-destruct & \textsc{selfdestruct}  \\
    & Low-level Calls & \textsc{call}, \textsc{callcode}, \textsc{delegatecall} \\ 
    \hline
    Payable & Transaction in Assets & \textsc{callvalue} \\
    \hline
    Pure & Gas Consumption & \textsc{stop}, \textsc{return}, \textsc{reverse} \\
    \bottomrule
  \end{tabular}
\end{table}

The \textit{view} function is not permitted to alter the state variable, so we are supposed to identify statements that can modify the state variable. In Table \ref{tab:statesOpcde}, we focus on storage modification, event emission, child contract creation, self-destruct, and low-level calls at the Opcode level. \textsc{sstore} first reads the \textit{key} and \textit{value} from the stack and then writes the \textit{value} at the \textit{key} address, which might overwrite the storage, modifying the state variables of contracts. Moreover, the contract's transaction information is saved in the state variable. When an event is invoked, the arguments are written to the transaction log, causing a change in the states. Thus, \textsc{log0}, \textsc{log1}, \textsc{log2}, \textsc{log3}, \textsc{log4} should be focused when emitting an event with different topics, even if \textsc{log}s do not affect the states. Additionally, creating or suiciding contracts will add or remove the storage and code of the contract, which is saved in the state variable. Hence, \textsc{create}, \textsc{create2}, and \textsc{selfdestruct} are identified as modifying the state variable. Furthermore, low-level calls (e.g., \textsc{call}, \textsc{callcode}, and \textsc{delegatecall}) are not permitted in the \textit{view} function, and the \textsc{staticcall} Opcode is used to replace these calls, which prohibits states modification. 

The keyword \textit{payable} is mandatory in any function that involves asset transactions, where the type and amount of the asset depend on the message. After the Solidity 0.5.2 version, the \textsc{callvalue} is used to obtain the value of the call.

The \textit{pure} function cannot read or modify states. To ensure that the function consumes no gas, we identify all Opcodes that are disallowed in \textit{pure} functions based on Ethereum's gas calculation. We exclude stack operations such as \textsc{push}, \textsc{pop}, \textsc{swap}, and \textsc{dup}, leaving only \textsc{stop}, \textsc{return}, and \textsc{reverse} as valid instructions for zero gas consumption \cite{wood2014ethereum}.
% according to the Ethereum Yellow Paper 

\subsection{Vulnerability Detection}
\label{subsec:vulnerability_detection}

\begin{figure*}[ht]
  \centering
  \includegraphics[width=\linewidth]{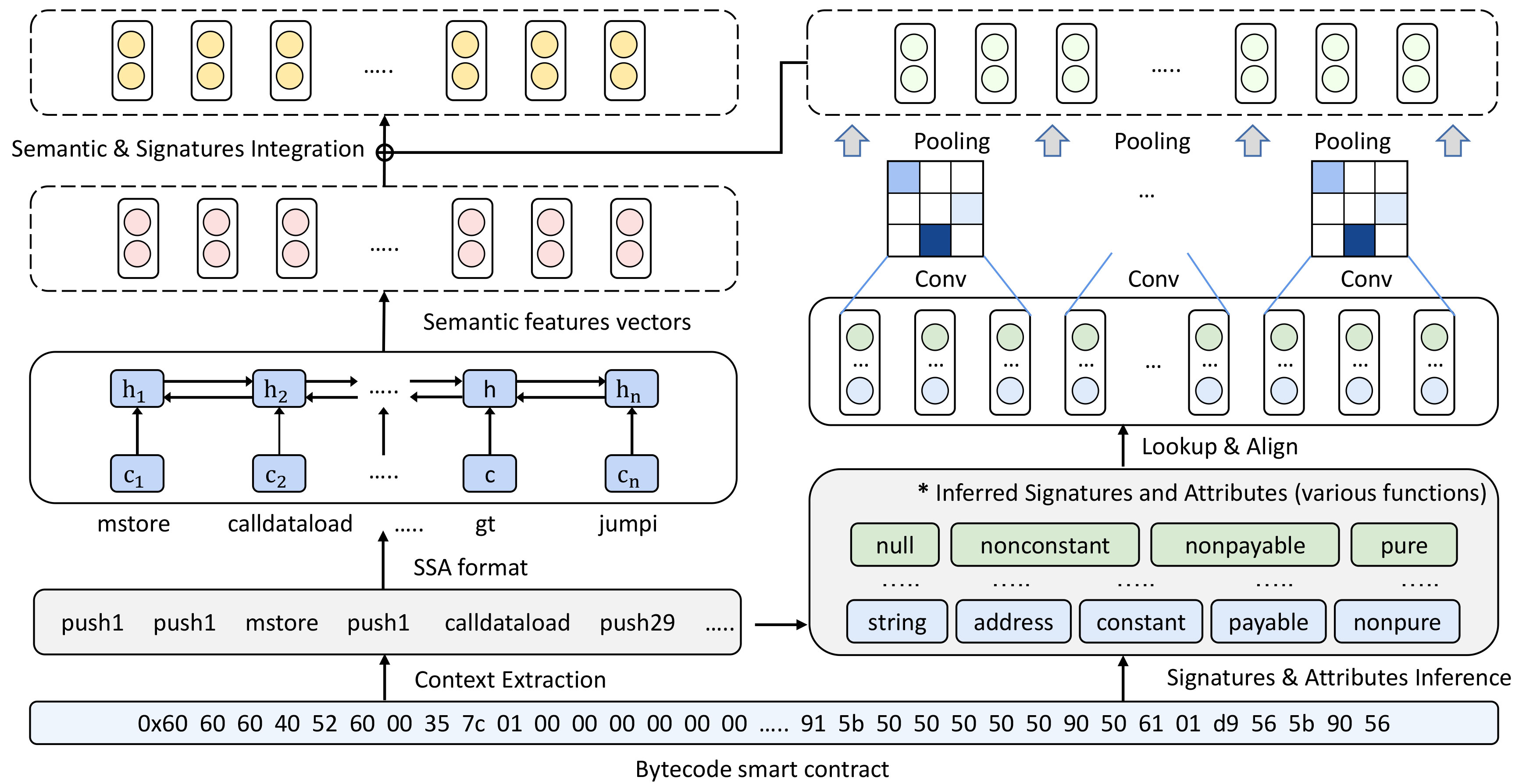}
  \caption{The Encoder of our Detection Model. The bottom layer is the hex 2-gram representation of the bytecode smart contract, which is converted into two types of data: the Opcode sequence and the function data. * indicates the part that can be replaced with raw ABI information. The box with yellow dots at the top of the diagram reflects the representation in the latent space.}
  \label{fig:model_detection_encoder}
  % \vspace{-2ex}
\end{figure*}

In this subsection, the main structure of our detection model will be presented. As Figure \ref{fig:model_detection_encoder} shows, the raw bytecode smart contract will first be processed by context extraction (\cref{subsec:contextExtraction}), signature inference (\cref{subsec:sigInference}) and attributes summarization (\cref{subsec:attrInference}). The sequence $X=\{x_1, x_2, ..., x_n\}$ contains $n$ Opcodes. Using the rattle tool \cite{crytic2020rattle}, we convert $X$ to a SSA format $S=\{s_1,s_2, ..., s_m\}$. Since the stack operations are eliminated, and the Opcodes are arranged in the order of execution, more precise semantic information can be obtained. One-hot encoding is utilized to convert $S$ into a $k$-dimensional embedding, and then feed a Bi-GRU layer to obtain the representations of the contract semantics in latent space, $h(j)=BiGRU(h_{j-1}, s_j)$, where $0<j\leq k$. In Figure \ref{fig:model_detection_encoder}, the inferred function signature can be replaced with ABI information. When the ABI data is retrieved, it is converted into machine-readable form by looking up the vocabulary. Convolutional neural network (CNN) and average pooling layer transform the ABI into a function feature representation, before concatenating semantic and function representations to form the final feature representation of the contract. 
% If the raw ABI data cannot be retrieved, all the details of the encoder will be described subsequently.

If ABI cannot be obtained, we employ the method in \cref{subsec:sigInference} to get the contract parameters for each function. Variation in the dimension of function signatures is inevitable, given multiple functions in each contract. Concurrently, the number of arguments varies, necessitating the alignment of sequences with varying lengths.
Assume that each function parameter prediction process has $z$ associated parameter types $T=\{t_1,t_2, ..., t_z\}$, and let $F_i$ represent the $i$-th sequence of functions ordered by the first basic block address $\{type^i_1,type^i_2,...,type^i_b\}$, containing a total of $b$ arguments. Finally, we concatenate the sequence of parameters $\{attr^i_1, attr^i_2, attr^i_3\}$ for $ap$ functions, where $ap$ is the number of functions in the contract. In order to get a representation of the function signatures at the contract level, we cluster the functions of each contract and then feed a CNN and average pooling layer. The factor that selects CNN is to obtain local feature vectors for signature data, and its convenience is another factor in our task. Specifically, we use a lookup dict to convert each parameter to a uniform numeric format and blank padding to align these parameters from various functions.

\begin{figure}[ht]
  \centering
  \includegraphics[width=0.8\linewidth]{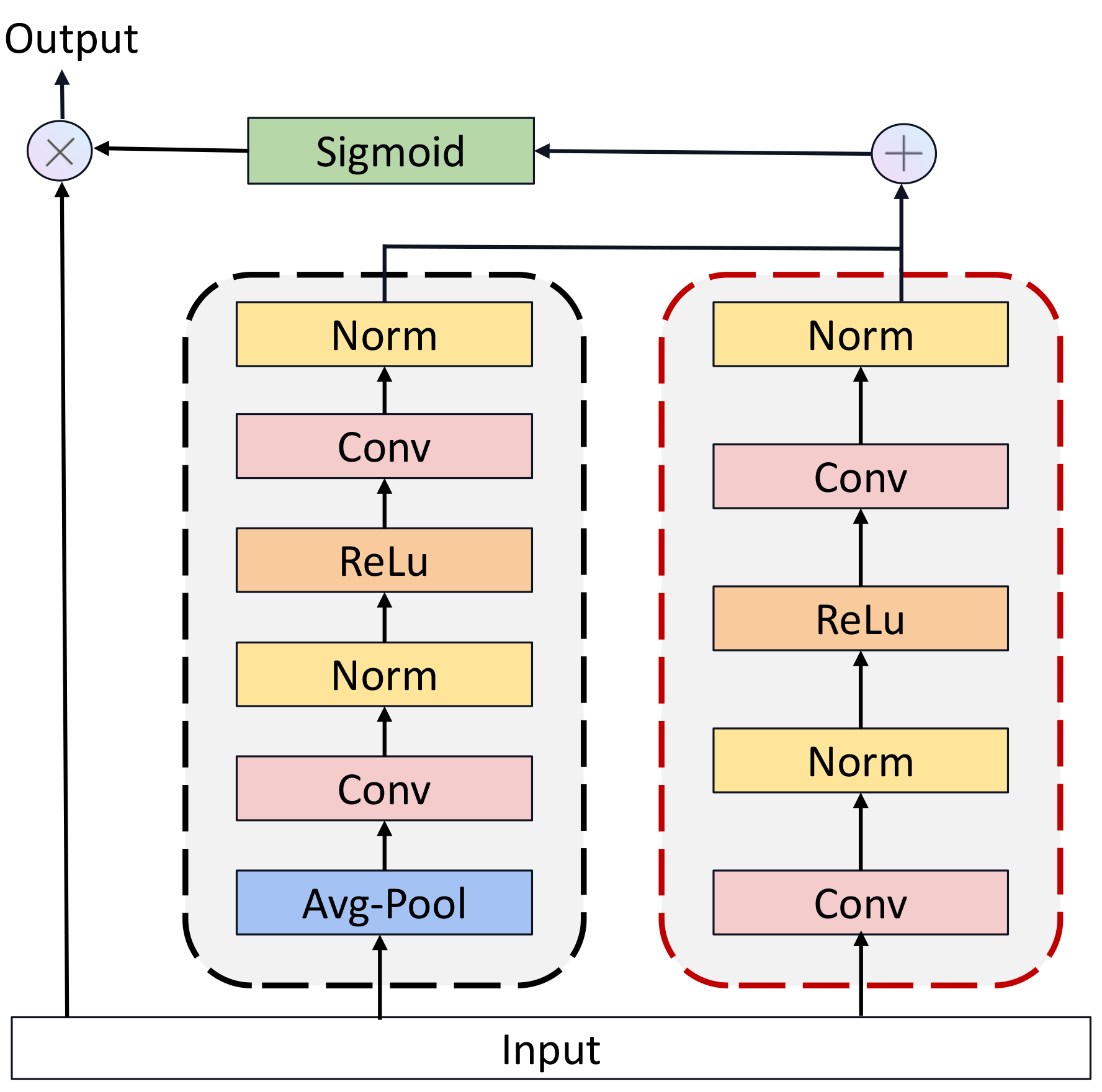}
  \caption{The MS-CAM in Our Structure. $\otimes$ means multiplication, $\oplus$ denotes addition. The black dashed box represents the global feature, and the red dashed box represents the local feature.}
  \label{fig:ms-cam}
\end{figure}

Supposing that $g_{ij} = type^i_j$ is the $i$-th function of $j$-th parameter in contract $X$, where $0 < i \leq ap, 0 < j \leq n$. We first feed them to several convolutional layers to get the features of each function, ${G'_i} = \boldsymbol{W_1}[g_{i1}, g_{i2}, ..., g_{i(1+k-1)}; $${attr^i_1},$ ${attr^i_2},$ ${attr^i_3}]$ $+ \boldsymbol{b_1}$ where the $\boldsymbol{W_1}, \boldsymbol{b_1}$ are the training parameters and $k$ represents the kernel size. To obtain the hidden features among all the functions, \textsc{MS-CAM} \cite{dai2021attentional} in Figure \ref{fig:ms-cam} is applied to obtain the features in the certain dimension $d_1$. \textsc{MS-CAM} was originally proposed for integrating features with different dimensions. We focus on the local features expressed by a certain function and the global features expressed by all functions. In this way, we can capture the related features of the individual functions. The local feature representation $feature_l$ can be expressed as the following \cref{eqa:MS-CAM_local}.

\begin{equation}
\begin{aligned}
    feature_l = N(Conv_2(\zeta(N(Conv_1({G})))))
\label{eqa:MS-CAM_local}
\end{aligned}
\end{equation}
where $G = \{g_1,g_2...,g_{ap}\}$, $N$ denotes the normalization layer, $Conv_1$ denotes shrinking the input sizes on dimension $d_1$, while $Conv_2$ denotes expanding the size on $d_1$ back to its original size. $\zeta$ represents the $ReLu$ activation function. The global feature $feature_g$ and output are calculated as the \cref{eqa:MS-CAM_global}, \cref{eqa:MS-CAM_output}.

\begin{equation}
\begin{aligned}
    feature_g = GAP(feature_l)
\label{eqa:MS-CAM_global}
\end{aligned}
\end{equation}
\vspace{-2ex}
\begin{equation}
\begin{aligned}
    output = {G}\otimes(\xi(feature_g \oplus feature_l))
\label{eqa:MS-CAM_output}
\end{aligned}
\end{equation}
where the $GAP$ represents a global average pooling layer, $\otimes$ means multiplication in the feature map, which is consistent with the \textsc{MS-CAM}. $\oplus$ denotes addition after adjustment, while broadcasting in \textsc{MS-CAM}. $\xi$ represents the $Sigmoid$ activation function.

Then, a $ReLu$ and a pooling layer are utilized to get the feature representation of the function parameters and attributes, $G^*_i = ReLU( G'_i )$, $G = Pool( G^*_1 , G^*_2 , ..., G^*_{ap} )$. Thus, the final feature representation of the contract $X$ is $h_x = [h(1), ... ,h(k); G]$.

Following our model's encoder, we obtain an implicit feature representation $h_x$ incorporating contract semantic and function signature information. To demonstrate the efficacy of our encoder, we adopt the same architecture as the decoder of \textsc{SRIF} model in \cref{subsec:sigInference}. We employ an attention structure for $h_x$, and a recurrent neural network is fed to decode the hidden states and identify the vulnerabilities. Moreover, since there is no same label in each prediction, a mask mechanism is used to get distinct results at different time steps. For example, if $y_j$ has the highest probability at time step $t-1$, the initial value of a label $y_j$ is set to negative infinity at time $t$, and set the initial value of all labels except $y_j$ to 0. To alleviate the class imbalance problem in the multi-class detection task, we use the focol loss function in the vulnerability detection process, which has been introduced in \cref{eqa:focalLoss_fineT} in \cref{subsec:sigInference}.

\section{Experiments}
\label{sec:experiments}
In this section, we present the results of experiments to evaluate the performance of our framework, answering the following questions:

\noindent RQ1: \textbf{ Why the RNN is chosen for \textsc{srif} and \textsc{cobra}?}\\
\noindent RQ2: \textbf{ Is \textsc{srif} effective for the function signature inference?}\\
\noindent RQ3: \textbf{ Will \textsc{srif} be affected by different compiler versions?}
\noindent RQ4: \textbf{ Is \textsc{cobra} effective for the vulnerability detection?}\\
\noindent RQ5: \textbf{ Is ABI or function signature in \textsc{cobra} effective?}\\
% How effective is \textsc{COBRA} with ABI and function signature separately?\\
% How do the ABIs or function signatures influence \textsc{cobra}?\\
\noindent RQ6: \textbf{ Can \textsc{cobra} detect new bugs in the real world?}

\subsection{Experimental Setup}
\label{experiment:setup}

\subsubsection{\textbf{Datasets}}
\label{sec:subsub_datasets}
To generate a labeled bytecode smart contract dataset \uppercase\expandafter{\romannumeral1} with a sufficient amount of ground truth for proper evaluation, we collect 13,948 deployed bytecode smart contracts from the \textsc{XBlock\_Eth} dataset \cite{zheng2020xblock}. We use the \textsc{SmartBugs} \cite{durieux2020empirical} framework to identify vulnerabilities in these contracts and label them accordingly. Due to the fact that each part of the tool has its own specialized vulnerabilities, we utilize various state-of-the-art modules to collect as much accurate ground truth data as possible. For instance, we utilize the \textsc{Oyente} to identify reentrancy, and \textsc{Mythril} is maintained for arithmetic, unchecked low-level calls, and transaction ordering dependency vulnerabilities detection. Especially, the time manipulation is labeled by the \textsc{Conkas} in \textsc{SmartBugs}, which is renewed by the community. In addition, we run each contract for a minimum of 30 minutes to ensure maximum reliability. After filtering out contracts that can not be detected due to version incompatibility and disassembly errors, we have obtained 8,267 processed contracts with their corresponding vulnerability labels.

Table \ref{tab:label_distribution} presents our dataset composition, comprising 790 analyzed contracts. Among these, we identified 790 instances of reentrancy vulnerabilities, 4,609 instances of arithmetic vulnerabilities, 1,764 cases of unchecked low-level calls, 1,351 transaction order dependencies, and 1,292 timing manipulation issues. Notably, 7,190 contracts were found to be vulnerability-free. The total amount of vulnerability exceeds the total number of contracts because most vulnerable contracts contain multiple different types of vulnerability.

\begin{table}[ht]
\small
\centering
  \caption{The Label Distribution of the Dataset I.}
  \label{tab:label_distribution}
  \begin{tabular}{c c}
    \toprule 
    \textbf{Vulnerability Types} & \textbf{Existing Amount} \\
    \midrule
    Reentrancy Vulnerability        &  790 \\
    Arithmetic Vulnerability        &  4,609 \\
    Unchecked Low Level Calls       &  1,764 \\
    Transaction Ordering Dependency &  1,351 \\
    Time Manipulation               &  1,292 \\
    No Vulnerability                &  7,190 \\
    \bottomrule
  \end{tabular}
  \vspace{-2ex}
\end{table}

Another dataset \uppercase\expandafter{\romannumeral2}, also derived from \textsc{XBlock\_Eth}, contains only contracts in bytecode format without vulnerability labels. We collected a total of 6,024 contracts for the phase of function signature inference. Using the function Opcodes and hashes acquisition method described in \cref{subsec:contextExtraction}, we gathered the function ids present in these contracts. These function signatures were then matched against the 4byte \cite{4byte} database. Finally, 99,745 function signatures, along with the corresponding Opcodes, were collected.

\subsubsection{\textbf{Evaluation Metrics}} We use F1-score, precision, and recall as evaluation metrics. The precision is the likelihood of each classification being accurately identified. Recall indicates the probability of discovering all possible results. The F1-score represents the harmonic mean of precision and recall.

% When calling a function, distinct function arguments can result in different function calls. To determine whether the function signature is accurate, it is necessary to obtain information about each function argument. 

\subsubsection{\textbf{Environments}}
Two experimental environments exist in the whole experiments, (1) Intel(R) Xeon(R) W-2255 CPU + 256GB RAM + 2 $\times$ GeForce RTX 3090 with the operating system of Windows Server 2019 and (2) Intel(R) Core(TM) I7-12700 CPU and 32GB RAM with the system of Ubuntu 20.04. In (1), we labeled contracts with vulnerability classes and trained models; in (2), we collected all bytecode smart contracts used in our framework from the \textsc{XBlock\_Eth}.

\subsection{RQ1:  Why is the RNN chosen for \textsc{srif} and \textsc{cobra}?}\label{subsec:rnn_choice}
\noindent\textbf{Motivation:} The reason why we choose the LSTM rather than other models (e.g., Transformer, and BERT) is (1) computational resource efficiency, and (2) robust handling of variable-length sequences. 

\noindent\textbf{Approach:} To comparatively assess model utilization efficiency across identical data conditions, we evaluate neuron coverage performance on the dataset described in \cref{sec:subsub_datasets}. Our experiments involve training 3 distinct architectures, i.e., Transformer, BERT, and LSTM, on dataset II, followed by coverage computation using the designated test partition. The dataset is divided following a 60\% : 40\% training and test ratio to ensure consistent evaluation conditions. Note that the conventional notion of a "neuron" requires careful interpretation in RNNs, as these architectures generate vector-based hidden states rather than discrete neuronal outputs~\cite{huang2021coverage}. Thus, we conduct coverage analysis by examining the hidden state vector at the testing layer. For Transformer and BERT, we follow \cite{wang2025tmf} setting the learning rate $l$ = 0.0001, hidden size $h$ = 128, batch size $\beta$ = 64, the number of attention heads $heads$ = 4, and the number of Transformer layers $L$ = 4. We set the threshold $\theta$ = 0 in the neuron coverage, which means that all the neurons whose parameter value is not 0 are valid.

\begin{figure}[ht]
  \centering
  \includegraphics[width=\linewidth]{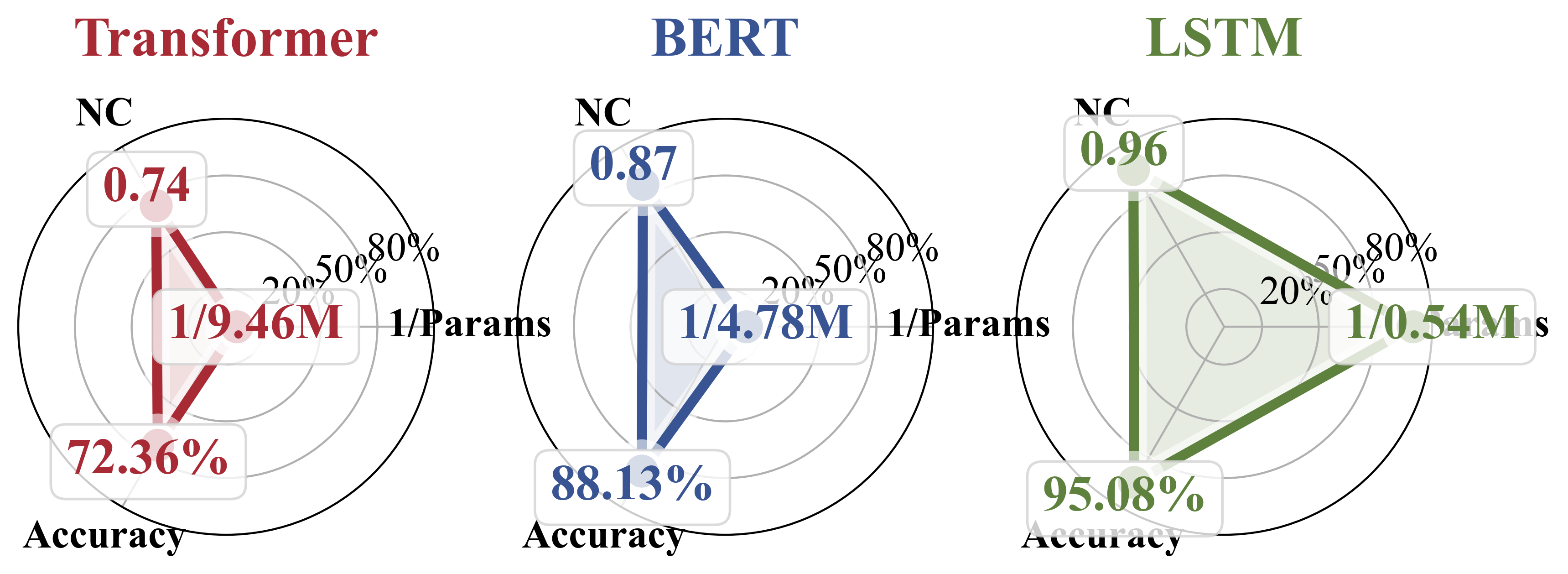}
  \caption{The Neuron Coverage Comparison of Different Model Types. The \textcolor[RGB]{167, 42, 53}{red color} represents the experimental data of the Transformer architecture, the \textcolor[RGB]{56, 84, 146}{blue color} represents the experimental data of the BERT architecture, and the \textcolor[RGB]{94, 129, 62}{green color} represents the experimental data of the LSTM. The three dimensions NC, Accuracy, and 1/Params are neuron coverage, accuracy, and the inverse of parameter amount, respectively.}
  \label{fig:NC_comparision}
% \vspace{-2ex}
\end{figure}

\noindent\textbf{Result:} Figure \ref{fig:NC_comparision} demonstrates the performance variations across Transformer, BERT, and LSTM when evaluated along three critical dimensions: neuron coverage (NC), detection accuracy (Accuracy), and parameter efficiency (i.e., 1/Params). LSTM achieves superior performance with 0.96 neuron coverage and 95.08\% detection accuracy, outperforming BERT's 88.13\% by 6.95 percentage points and Transformer's 72.36\% by 22.72 percentage points. Furthermore, LSTM maintains the performance advantage while demonstrating significantly greater parameter efficiency 1/0.54M, exceeding BERT's efficiency 1/4.78M by over 8 times and Transformer's 1/9.46M by around 17.5 times. The experimental data shows that LSTM can achieve better detection results with fewer model parameters in our dataset and tasks.

% \begin{table}[ht]
% \small
%   \caption{The Neuron Coverage Comparison of Different Model Types.}
%   \label{tab:NC_comparision}
%   \begin{tabular}{c c c c}
%     \toprule 
%     \textbf{Model Names} & \textbf{Parameter Amount} & \textbf{NC} & \textbf{Accuracy} \\
%     \midrule
%         Transformer & 9.46 Million  &  0.74      &  72.36\%   \\
%         BERT        & 4.78 Million  &  0.87      &  88.13\%   \\
%         LSTM        & 0.54 Million  &  0.96      &  95.08\%   \\
%     \bottomrule
%   \end{tabular}
%   % \vspace{-2ex}
% \end{table}

\begin{tcolorbox}[boxrule=1pt,boxsep=1pt,left=3pt,right=3pt,top=3pt,bottom=3pt]
\textbf{Answer to RQ1.}
LSTM demonstrates superior neuron coverage, detection accuracy, and parameter efficiency relative to Transformer and BERT architectures.
\end{tcolorbox}

\subsection{RQ2: Is \textsc{srif} effective for the function signature inference?}\label{subsec:rq1}
\noindent\textbf{Motivation:} We first verify the effectiveness of SRIF on function signature data. It serves as the foundational element for ensuring the efficacy of COBRA.

\noindent\textbf{Approach:} To evaluate the effectiveness of various network structures, we divide the dataset \uppercase\expandafter{\romannumeral2} into training, validation, and test sets with proportions of 60\%, 20\%, and 20\%, respectively. After training, performance results from various model structures are collected during validation, and the best network structure is evaluated on the test set.
% In this section, we present the results of our function signature inference experiments, answering the \textbf{RQ1}.
Due to the fact that each function call procedure is composed of distinct basic blocks, the flow between each block is uncertain and diverse, resulting in multiple branches. Therefore, we employ the DFS algorithm to obtain the Opcode of each function in the executing flow. To determine the optimal depth, we compare the instances of 1, 2, and 3 depths, respectively. Furthermore, we compare the SRIF with Gigahorse~\cite{grech2019gigahorse} on a subset of the test set. The contracts are compiled manually in the Gigahorse tool and their reverse recovered function signature results are collected. It is worth noting that at this stage, we strictly control the number and order of function parameters, and when the number and order are inconsistent, we consider that the function signature decision fails.

\noindent\textbf{Result:} We compare the \textsc{LSTM}, \textsc{GRU} cells, and different depths in SRIF. According to the results in Table \ref{tab:inferParamResult}, the ideal results are obtained when the depth is 1. In light of this result, we presume that the information most relevant to function parameters is stored in the first basic block of the function, which is the location of the function entry. The results indicate that the \textsc{LSTM} owns 95.46\% F1-score, which has a more favorable performance and is better suited for the stage of function inference.

\begin{table}[ht]
\small
  \caption{The Results of Function Parameter Inference in Different Depths and Cells with Focal Loss Function}
  \label{tab:inferParamResult}
  \begin{tabular}{c c c c c}
    \toprule 
    \multicolumn{2}{c}{\textbf{Network Structures}} & & & \\
    \cline{1-2}    
    \textbf{Cells} & \textbf{Max Depths} & \textbf{F1-socre} & \textbf{Precision} & \textbf{Recall} \\
    \midrule
    GRU & 3 & 93.19\% & 90.51\%  & 96.04\% \\
    LSTM & 3 & 95.23\% & 94.14\%  & 96.36\% \\
    GRU & 2 & 94.85\% & 93.75\%  & 95.96\% \\
    LSTM & 2 & 95.20\% & 93.99\%  & 96.44\% \\
    GRU & 1 & 95.18\% & 94.16\%  & 96.23\% \\
    \textbf{LSTM} & \textbf{1} & \textbf{95.46\%} & \textbf{94.17\%}  & \textbf{96.78\%} \\
    \bottomrule
  \end{tabular}
\end{table}

Furthermore, we use the cross entropy loss function for training, obtaining the F1-score, precision, and recall rate of 94.46\%, 93.81\%, and 95.62\%, respectively. Note that the cell and depth are \textsc{LSTM} and 1 separately. The result reveals that the focal loss function has a certain improvement effect than the cross entropy loss function.

% Then, to improve the model's capacity for predicting function signatures, we compare the \textsc{LSTM} and \textsc{GRU} cells in our model. The results indicate that the \textsc{LSTM} owns 95.46\% F1-score, which has a more favorable performance and is better suited for the stage of function inference. Furthermore, to reveal that the focal loss function is available, we use the cross entropy loss function for training to obtain F1-score, precision, and recall rate of 94.46\%, 93.81\%, and 95.62\%, respectively. Note that the cell and depth are \textsc{LSTM} and 1 separately. This experimental result reveals that the focal loss function has a certain improvement effect.

The model for function parameters inference consists of 540,771 parameters when \textsc{LSTM} and focal loss function is employed. As Table \ref{tab:test_bytecode2signature} shows, \textsc{SRIF} can achieve 94.76\% F1-score, 93.49\% precision, and 96.06\% recall, which indicates that it can achieve high performance in function signature inference of smart contracts.

\begin{table}[ht]
\small
  \caption{The Measures of Function Parameters Inference.}
  \label{tab:test_bytecode2signature}
  \begin{tabular}{p{4cm} p{3.5cm}}
    \toprule 
    \textbf{Metrics for Testing} & \textbf{SRIF Performance} \\
    \midrule
        Test Precision  & 93.49\% \\
        Test Recall  &  96.06\% \\
        Test F1-Score & 94.76\% \\
    \bottomrule
  \end{tabular}
\end{table}

Moreover, we randomly select 12 contracts (a total containing 120 functions) in the test set, comparing the SRIF with the Gigahorse. Due to the problem of time consumption, only part of the contracts are selected in this paper. During the selection, all the contracts in the test set are randomly shuffled and divided into 12 equidistant intervals, and the contracts are randomly selected from each interval. This sufficient randomness gives some validity to the results. In this process, we find that Gigahorse can successfully recover 98 function signatures, but SRIF successfully recovers 110 function signatures.

\begin{tcolorbox}[boxrule=1pt,boxsep=1pt,left=3pt,right=3pt,top=3pt,bottom=3pt]
\textbf{Answer to RQ2.}
The SRIF achieves 94.76\% F1-score on the test set, and it can recover more function signatures than Gigahorse.
\end{tcolorbox}

\subsection{RQ3:  Will \textsc{srif} be affected by different compiler versions?}

Our evaluation of \textsc{srif}'s effectiveness utilizes a comprehensive dataset of unoptimized, open-source smart contracts compiled across multiple Solidity versions. The dataset construction leverages two key resources: (1) version metadata from Etherscan \cite{etherscan2023io} and (2) compilation verification through the official EVM Solc compiler. From the initial two million blockchain blocks, we extracted 13,948 reliability-focused contracts. These represent 85 distinct Solidity compiler versions spanning v0.4.11 through v0.8.30.

\begin{figure}[ht]
  \centering
  \includegraphics[width=\linewidth]{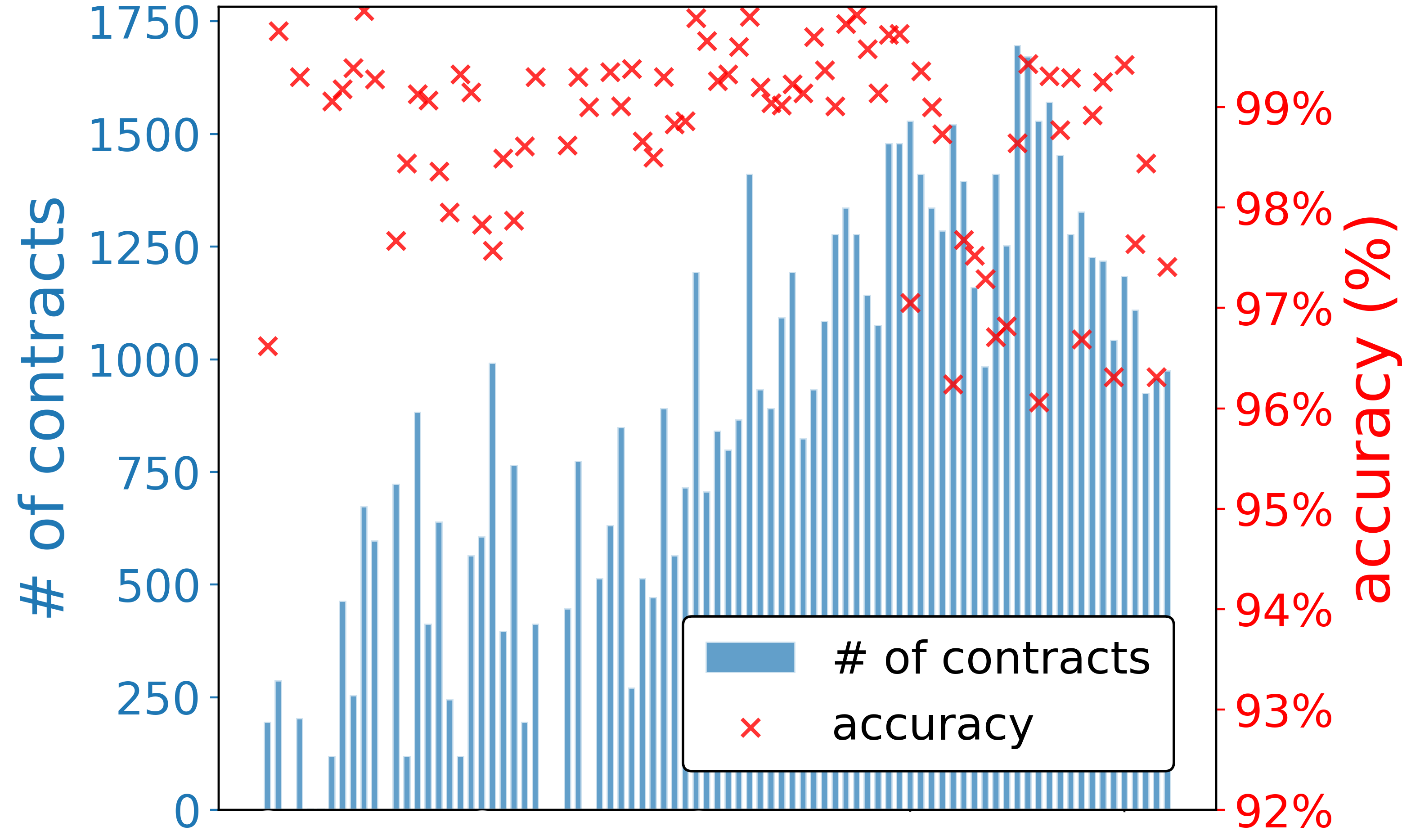}
  \caption{Performances of \textsc{srif} for Different Solidity Compiler Versions. The \textcolor[RGB]{31,119,180}{blue bars} represent the number of contracts under different compiler versions, and the \textcolor{red}{×} indicates the inferred accuracy under different compiler versions.}
  \label{fig:SRIF_compiler_versions}
\end{figure}

For each compiler version, we evaluate \textsc{srif}’s accuracy performance. Figure \ref{fig:SRIF_compiler_versions} presents these results, sorted by Solidity version in ascending order, along with the corresponding number of contracts per version, ranging from 1 to 1750.
Our experimental results demonstrate \textsc{srif}'s robust performance across all tested compiler versions. In the Solidity compilers, the tool maintains consistently high accuracy, with no observed case falling below 96\% across all 85 versions. This finding confirms that \textsc{srif}'s accuracy remains stable regardless of compiler version evolution.

\begin{tcolorbox}[boxrule=1pt,boxsep=1pt,left=3pt,right=3pt,top=3pt,bottom=3pt]
\textbf{Answer to RQ3.}
The evaluation results indicate that \textsc{srif} maintains stable performance across different compiler configurations. It achieves consistent accuracy, with $\geq$96\% success rate for all 85 Solidity compiler versions tested.
\end{tcolorbox}

\subsection{RQ4: Is \textsc{cobra} effective for the vulnerability detection?}
\noindent\textbf{Motivation:} We evaluate COBRA's ability to detect vulnerabilities in our collected dataset.

\noindent\textbf{Approach:} We utilize the dataset \uppercase\expandafter{\romannumeral1}, which contains 8,267 labeled contracts. We start by conducting comparisons among various LSTMs, GRU cells, loss functions, and other tools using the validation set. Subsequently, we assess COBRA's performance on the test set to derive the final evaluation results. Additionally, we explore a scenario where only inferred function signatures are accessible. The F1-score serves as the metric, representing the harmonic mean of precision and recall, thereby evaluating the COBRA, Mythril~\cite{durieux2020empirical}, and MANDO-GURU~\cite{nguyen2022fse}. The recall signifies the percentage of identified malicious classes among the actual malicious classes, evaluating when only inferred signatures are available.

\noindent\textbf{Result:} As Table \ref{tab:vd_results} shows, under the combination of context and ABI, the \textsc{GRU} with focal loss function can achieve the best F1-score (94.26\%) in our evaluation. Notably, our model can achieve even higher recall on cross-entropy. One of the possible reasons is that the focal loss function in our experiments might result in more training for the more challenging classes, reducing generalization for the simpler categories. Nevertheless, \textsc{GRU} can perform better than \textsc{LSTM}. Furthermore, after we compare Mythril and MANDO-GURU with COBRA, we find that COBRA can achieve the best performance as shown in Table \ref{tab:comparison_COBRA}.

\begin{table}[ht]
\small
  \caption{The Performance Comparison of Different Cells and Loss Functions.}
  \label{tab:vd_results}
  \begin{tabular}{c c c c c}
    \toprule 
    \multicolumn{2}{c}{\textbf{Network Structures}} &  & & \\
    \cline{1-2}
    \textbf{Cells} & \textbf{Loss Function} & \textbf{F1-socre} & \textbf{Precision} & \textbf{Recall}\\
    \midrule
        LSTM & Cross Entropy & 92.38\% & 90.82\%  & 94.01\% \\
        LSTM& Focal Loss & 87.96\% & 84.82\% & 90.49\% \\
        GRU & Cross Entropy & 93.17\% & 90.40\% & 96.13\% \\
        \textbf{GRU} & \textbf{Focal Loss} & \textbf{94.26\%} & 93.12\% & 95.42\%\\ 
    \bottomrule
  \end{tabular}
  % \vspace{-3ex}
\end{table}

\begin{table}[ht]
\small
  \caption{The Comparison Results of COBRA and SOTA.}
  \label{tab:comparison_COBRA}
  \begin{tabular}{p{3.5cm} p{3cm}}
    \toprule 
    \textbf{Name} & \textbf{F1-score} \\
    \midrule
        Mythril & 65.57\% \\
        MANDO-GURU  & 91.06\% \\
        COBRA  &  94.26\% \\
    \bottomrule
  \end{tabular}
  \vspace{-2ex}
\end{table}

When combining context, the global features of ABI with GRU, and the focal loss function, the model contains a total of 3,130,115 parameters. In the test phase, the results in Table \ref{tab:test_detections} show that 93.45\% of the F1-score, 91.56\% of the precision, and 95.42\% of the recall score can be obtained. Moreover, we utilize the \textsc{srif} with attributes summarization method to generate alternative function information for contracts that do not expose ABI. As a reminder, the method needs to support the ability to discover as many vulnerabilities as possible. Therefore, we take recall as the metric for detection. After testing, Table \ref{tab:test_detections} shows the recall can reach 89.46\%.

\begin{table}[ht]
\small
  \caption{The Measures of Vulnerabilities Classification.}
  \vspace{-1ex}
  \label{tab:test_detections}
  \begin{tabular}{p{3.5cm} p{3cm}}
    \toprule 
    \textbf{Metrics for Testing} & \textbf{COBRA Performance} \\
    \midrule
        Test Reminder Recall & 89.46\% \\
        Test Precision  & 91.56\% \\
        Test Recall  &  95.42\% \\
        Test F1-Score &  93.45\% \\
    \bottomrule
  \end{tabular}
  \vspace{-2ex}
\end{table}

\begin{tcolorbox}[boxrule=1pt,boxsep=1pt,left=3pt,right=3pt,top=3pt,bottom=3pt]
\textbf{Answer to RQ4.}
The COBRA achieves 93.45\% F1-score on the test set, and 94.26\% F1-score on the validation, which outperforms other methods.
\end{tcolorbox}

% To evaluate the effectiveness of our approach, we compared SOTA tools on a public dataset in \cite{liu2021smart}. As this dataset lacks address information, the Table \ref{tab:performance_comparison} displays the results of the detection with the inferred information. 

% \begin{table}[ht]
%   \caption{The Performance Comparison.}
%     \small
%   \label{tab:performance_comparison}
%   \begin{tabular}{c c c c c}
%     \toprule 
%     \textbf{Vuln. kind}& \textbf{Num.} & \textbf{Reminder} & \textbf{Mythril} & \textbf{} \\
%     \midrule
%        Reentrancy  &  146  & 140 &  52     & \\ 
%        Arithmetic   &  70  & 66  &  25  & \\
%        Time Manipulation & 77 & 76 &  58 & \\
%     \bottomrule
%   \end{tabular}
% \end{table}

\subsection{RQ5: Is ABI or function signature in \textsc{cobra} effective?}

\noindent\textbf{Motivation:} We conduct experiments to explore scenarios where different function interfaces are available, i.e., application binary interface (ABI) or function signatures.

% We verify the impact of the function interface in \textsc{cobra}. We evaluate the performance of \textsc{cobra} with ABI and function signature, respectively. 
% Finally, we test to demonstrate our capability for detection.

\noindent\textbf{Approach:} To demonstrate that ABI information is valuable for vulnerability detection, we summarize the case of distinct contract semantics and the addition of ABI separately. The experiments involved with ABI are conducted by the combination structure of \textsc{LSTM} with cross-entropy loss function. Furthermore, regarding the function signatures, we make a comparison of different network structures.  When no public ABI is available, inferred function signature representations and semantic features are used as latent features of contracts. Since we expect the COBRA with only function signatures to discover as many malicious classes as possible, recall is utilized to evaluate this situation.

\begin{table}[ht]
\small
  \caption{The Performance Comparison of Different Composite Structures.}
  \vspace{-1ex}
  \label{tab:feature_result}
  \begin{tabular}{c c c c c}
    \toprule 
    \textbf{Network Structures} & \textbf{F1-socre} & \textbf{Precision} & \textbf{Recall} \\
    \midrule
        Context          & 76.27\% & 70.21\%  & 83.47\% \\
        Context + ABI    & \textbf{92.38\%} & \textbf{90.82\%} & \textbf{94.01\%} \\
        Context + ABI + MS-CAM & \textbf{92.14\%} & \textbf{91.34\%} & \textbf{92.96\%} \\
    \bottomrule
  \end{tabular}
  \vspace{-2ex}
\end{table}

\noindent\textbf{Result:} Table \ref{tab:feature_result} summarizes the case of distinct contract semantics and the addition of ABI separately. Note that all the data in Table \ref{tab:feature_result} is done by the combination structure of \textsc{LSTM} with cross-entropy loss function. Table \ref{tab:feature_result} provides insight into numerous conclusions. First, general results can be obtained using only context or SSA Opcode. The relatively low precision indicates that the features cannot effectively specify the vulnerabilities. In addition, after combining ABI, we compared the different ABI features using the global feature and the whole \textsc{ms-cam}. The results presented indicate that using only global feature extraction can improve F1-score and recall, while \textsc{ms-cam} can enhance precision.

\begin{table}[ht]
\small
  \caption{The Comparison of Different Architectures for Vulnerability Detection with Inferred Function Signatures.}
  \vspace{-1ex}
  \label{tab:inferred_function_detection}
  \begin{tabular}{c c c c c c}
    \toprule 
     & \multicolumn{2}{c}{\textbf{LSTM}} & & \multicolumn{2}{c}{\textbf{GRU}} \\
    \cline{2-3}
    \cline{5-6}
    \textbf{Structures} & --- & \textbf{\textsc{ms-cam}}& & --- & \textbf{\textsc{ms-cam}} \\
    \midrule
       \textbf{Recall} & 87.79\%  &  89.01\% & & 89.78\% & \textbf{90.92\%}  \\ 
    \bottomrule
  \end{tabular}
  \vspace{-2ex}
\end{table}

As shown in Table \ref{tab:inferred_function_detection}, the \textsc{ms-cam} can obtain better performance when combined with GRU, and the recall value can increase to 90.92\%. Therefore, it is evident from the results in Table \ref{tab:inferred_function_detection} that the combination of local and global features of function features is more beneficial for vulnerability detection. The GRU and modified \textsc{ms-cam} can improve the detection. 
 % in the contract

\begin{tcolorbox}[boxrule=1pt,boxsep=1pt,left=3pt,right=3pt,top=2pt,bottom=2pt]
\textbf{Answer to RQ5.}
The ABI and function signatures can improve COBRA's detection performance to a certain degree.
\end{tcolorbox} 

\subsection{RQ6: Can \textsc{cobra} detect new bugs in the real world?}\label{subsec::rq4}
\noindent\textbf{Motivation:} We discuss the vulnerability detection capability of COBRA and analyze the vulnerability we detected.

\noindent\textbf{Approach:} We test in Xblock-ETH except for the contracts included in \cref{sec:subsub_datasets}, which are detected using COBRA. These contracts exist on the Ethereum mainnet. During this process, COBRA uses GRU and focal loss function. When the ABI is not publicly available, we add the MS-CAM module.

\noindent\textbf{Result:} We identify two previously undiscovered interactive vulnerabilities (CVE-2023-36979 and CVE-2023-36980), which can not be detected by Mythril, Oyente, and MANDO-GURU.
% \footnote{0x23a91059fdc9579a9fbd0edc5f2ea0bfdb70deb4}
An illustration of a potential contract vulnerability is presented in Listing \ref{lst:rq4}, which could be exploited. The contract depicts a casino game developed on the Ethereum blockchain. The global variable instance Casino has the uint balance, which ranges from $0$ to $2^{256}-1$. If the user needs to play the game, they need to store tokens into \texttt{casino.balance} according to the \texttt{casinoDeposit()} function.

\begin{lstlisting}[language=Solidity,
                    basicstyle=\footnotesize\tt,
                    numbers=left,
                    captionpos=b,
                    aboveskip = 2em,
                    belowskip = 1em,
                    numbersep = -1em,
                    caption= The Simplified Snippets of \textbf{Casino},
                    label=lst:rq4,
                    ] 
    function casinoDeposit() {
        if (msg.sender == casino.addr)
            casino.balance += msg.value;
        else 
            msg.sender.send(msg.value);
    }
    // Bet on Number
    function betOnNumber(uint number) public returns (string) {
        // Input Handling
        address addr = msg.sender;
        uint betSize = msg.value;
        if (betSize < casino.bettingLimitMin || betSize > casino.bettingLimitMax) {
            // Return Funds
            if (betSize >= 1*10**18)
                addr.send(betSize);
            return "Please choose an amount";
        }
        if (betSize * 36 > casino.balance) {
            // Return Funds
            addr.send(betSize);
            return "Casino has insufficient funds";
        }
        if (number < 0 || number > 36) {
            // Return Funds
            addr.send(betSize);
            return "Please choose a number";
        }
        // Roll the wheel
        privSeed += 1;
        uint rand = generateRand();
        if (number == rand) {
            uint winAmount = betSize * 36;
            casino.balance -= (winAmount - betSize);
            addr.send(winAmount);
            return "Win!";
        }
        else {
            casino.balance += betSize;
            return "Wrong number.";
        }
    }
\end{lstlisting} 

We can find that it only limits the situation of guessing correctly, and when it is correct, the money can be withdrawn. However, for the 35 lines of correct guessing at Listing \ref{lst:rq4}, the \texttt{send()} is used to transfer the money amount. If the transfer is not successful, it will only return false and will not block execution, so the return value needs to be checked. However, the owner does not check the return value in the contract. Moreover, If the gas exceeds 2300, the transaction will fail. But the \texttt{fallback()} function needs at least 2300. Therefore, if the attacker designs the contract to call \texttt{betOnNumber(uint number)} using the \texttt{call} method in an attack function, and then jumps to the \texttt{fallback} function when the call ends. It causes the gas to exceed 2300, causing the \texttt{send()} function to fail.

\noindent\textbf{Impact:} Based on CVSS (Common Vulnerability Scoring System) assessments~\cite{CVSS2025First}, the analyzed vulnerabilities are categorized as medium-risk severity. As for CVE-2023-36979, we will introduce the details in the \cref{subsec:casestudy}. It originates from improper authorization mechanisms in low-level call operations, enabling malicious manipulation of the contract's balance state variable, which subsequently disrupts expected payment distributions to users due to flawed contract logic. Meanwhile, CVE-2023-36980 in Listing \ref{lst:rq4} stems from insufficient validation checks on monetary-related parameters, permitting unauthorized alteration of transfer values that ultimately result in permanent financial losses for contract participants. Both security vulnerabilities demonstrate significant financial implications by exposing user assets to substantial risks of exploitation and irreversible damage.

\begin{tcolorbox}[boxrule=1pt,boxsep=1pt,left=3pt,right=3pt,top=3pt,bottom=3pt]
\textbf{Answer to RQ6.}
Utilizing COBRA, we find two previously undisclosed vulnerabilities, i.e., CVE-2023-36979 and CVE-2023-36980.
\end{tcolorbox}

\section{Discussion}
\label{sec::dis}

\subsection{Properties of \textit{COBRA}}\label{subsec:features}
COBRA achieves high performance by combining functional interfaces and contract semantics features. Compared to other deep learning technologies, i.e., MANDO-GURU, COBRA benefits from greater information features through its function interfaces.
In SRIF and COBRA, we use the RNN to achieve the sequential dependency of opcode execution. 

When learning the semantic features of a contract, both SRIF and COBRA employ a bidirectional RNN structure to apprehend the opcode sequence features stemming from the sequential execution of a contract. In addition, SRIF uses RNN architecture to sequentially decode the function interface parameters. Given the relatively limited variety of function parameters in Ethereum smart contracts, the label space during decoding is not excessively vast.

The deep learning architecture implemented in both COBRA and SRIF provides inherent extensibility for emerging signature and vulnerability patterns. When novel categories are identified, the system accommodates them through two straightforward modifications: (1) incorporation of additional training samples representing the new type, and (2) dynamic adjustment of the focal loss $\alpha$ parameter at \cref{eqa:focalLoss_fineT1} to maintain appropriate class weighting based on updated vulnerability frequency distributions.

\subsection{Runtime Overhead}\label{subsec:runtimeoverhead}
COBRA needs to obtain function interface information first, and then integrate it with contract semantics. 

When using ABI data exposure, this paper only needs to input it into the model at the same time as the contract code, which will not affect the time consumption of COBRA. 

When the ABI data is not public, through the experiments of \cref{subsec:rq1} in this paper, the time taken by the SRIF model to parse the function signature of a function is about 0.5667 seconds. This has a negligible impact on the time of the model in this paper. Moreover, in our dataset, the length of opcodes of several smart contracts exceeds 16,600. They are quite complex, leading to over 30 minutes of analysis by the symbolic execution tool Mythril. COBRA detected the entire test set only took a few minutes, performing relatively better. More importantly, the function-attribute interface summarization module can be determined to search for a fixed element from a specified sequence, even if the time complexity of the sequential search algorithm is linear $O(n)$, and does not consume too much time.

We compared the time consumption of Mythril, MANDO-GURU, and COBRA in the test set, but the results had a huge difference, so we did not intentionally record the data. We used Mythril to analyze the test set, which took several hours. In particular, when analyzing contracts with opcode lengths over 7300, it took almost ten minutes each. However, the COBRA detected the entire test set only took a few minutes (with fair CPU usage). It may be caused by the natural advantages of deep learning, making the timing too different. For the MANDO-GURU, a single contract takes several minutes (with fair CPU usage).

\subsection{Case Study}\label{subsec:casestudy}

\subsubsection{Arithmetic Vulnerability} 
\begin{figure}[ht]
  \centering
  \includegraphics[width=0.6\linewidth]{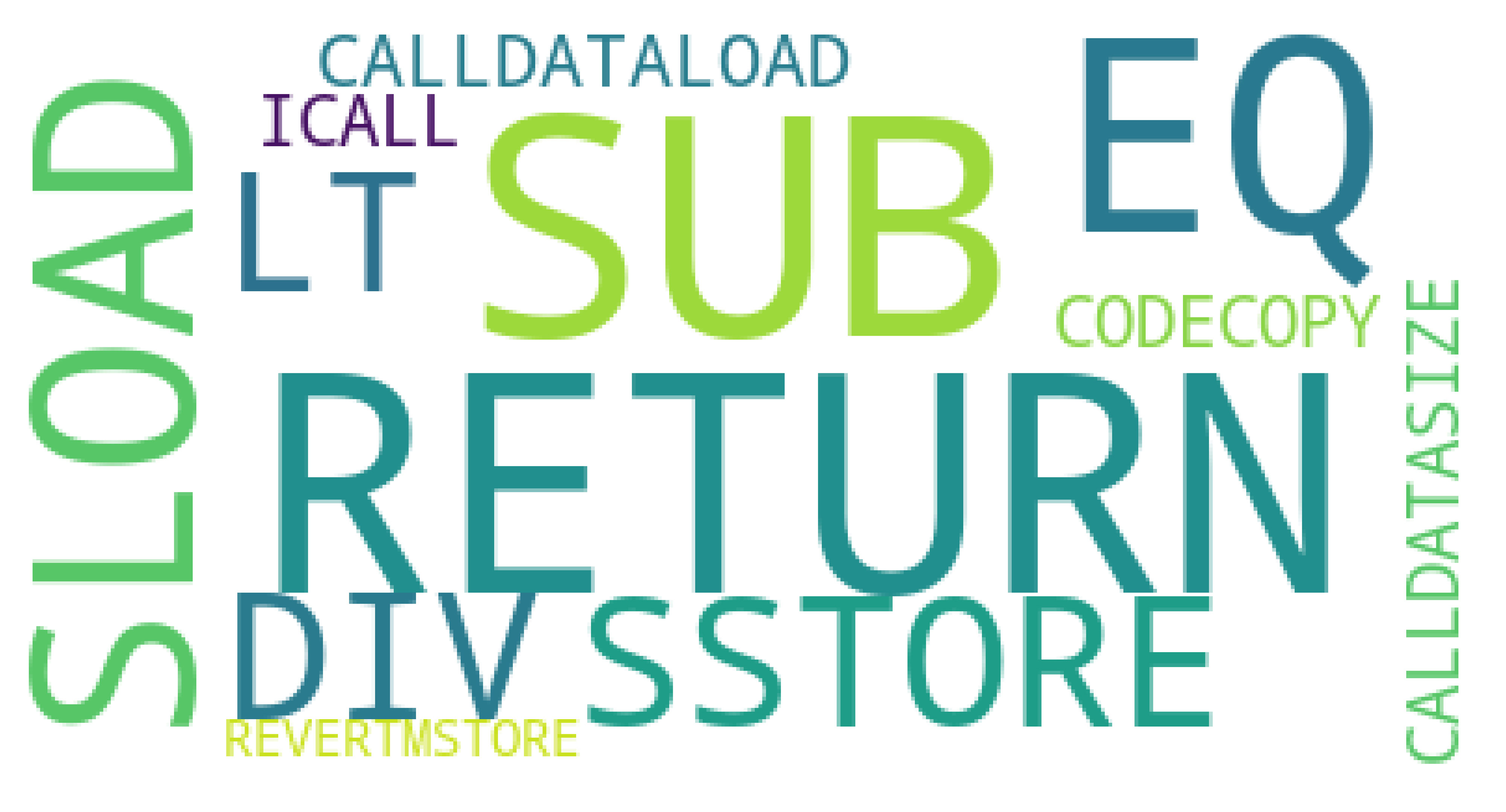}
  \caption{The Frequencies of SSA Opcodes in Arithmetic Vulnerabilities}
  \label{fig:arthmric_cloud_graph}
\end{figure}

In Figure \ref{fig:arthmric_cloud_graph}, we present the statistics of arithmetic vulnerabilities and summarize each contract's SSA Opcode by measuring dataset in \cite{liu2021smart}\cite{durieux2020empirical}.

To identify the Opcodes associated with the arithmetic vulnerability, we performed certain operations to eliminate the Opcodes (e.g., \textsc{calldataload}), which were almost present in any contract. We examined the frequency of each Opcode in each contract, dividing the total number of occurrences by the total number of contracts. To eliminate the effect of generic Opcodes, we select Opcodes with values close to 1, which can access data appearing in every weak contract. Figure \ref{fig:arthmric_cloud_graph} illustrates that numerous data operation instructions (e.g., \textsc{sub}, \textsc{eq}, \textsc{div}, and \textsc{lt}) exist. We also collect input parameters for these functions. Considering that our framework only detects integer overflows and underflows, we collect the function parameters and attributes containing integer types. Our function inference method demonstrates that 67.86\% of function contracts with integer-type parameters exist. In the meantime, these functions are not \textit{view} type, which has included instructions for modifying the state variables (e.g., \texttt{address.balance}, \texttt{block}, \texttt{tx}, and \texttt{msg}). 

\begin{lstlisting}[language=Solidity,
                    basicstyle=\footnotesize\tt,
                    numbers=left,
                    captionpos=b,
                    aboveskip = 1em,
                    belowskip = 1em,
                    numbersep = -1em,
                    caption= The Simplified Snippets of \textbf{Roulette},
                    label=lst:Integer,
                    ] 
  contract Roulette {
    uint ps; 
    struct Casino {
      address addr;
      uint balance;
    }
    Casino casino;
    function bet(uint num) public {
      address addr = msg.sender;
      uint betSize = msg.value;
      ps += 1;
      uint rand = genRand();
      uint randC = (rand + 1) % 2;
      if (rand != 0 && (randC == num)) {
        casino.balance -=  betSize * 2;
        addr.send(betSize * 2); }
      else { casino.balance += betSize; }
    }
    function genRand() private returns (uint) { 
      ps = ((ps * 3 + 1) / 2) % 10 ** 9;
      uint bn = block.number;  
      uint d = block.difficulty; 
      uint t = block.timestamp; 
      uint g = block.gaslimit; 
      uint rand = (ps + bn + d + t + g) % 37;
      return rand;
    }
  }
\end{lstlisting}

In Listing \ref{lst:Integer}, we give an example of the arithmetic vulnerability that we detected. Lines 11 and 17 of the code exhibit the potential for integer overflows, but we will focus on line 17 in this analysis. On line 3, a \textit{struct} with an \textit{address} and a \textit{uint} type is defined and stored in \textit{memory}. The variables \texttt{ps} and \texttt{balance} are unsigned integral numbers with a default range of $[0,2^{256}-1]$. The input parameter of the function \texttt{bet} is a \textit{uint}. When \texttt{balance} and \texttt{betsize} are out of range, such as when the \texttt{balance} is $2^{256}-1$, even a \texttt{betsize} of 1 can cause the \texttt{casino} instance to overflow, resulting in significant economic loss. Furthermore, the overflow in line 11 causes \texttt{ps} to become zero, and as a result, its random seeds can be deduced. As demonstrated in this example, it is challenging to identify whether an overflow has occurred by solely examining the SSA Opcode sequence. If the overflow in the function does not modify the state, it may result in a logic error without causing any direct economic loss. Therefore, our approach incorporates function parameters and function state types to determine whether a state operation occurs.

\subsubsection{Time Manipulation Vulnerability} We have also identified another type of vulnerability in the contract in Listing \ref{lst:Integer}, the time manipulation vulnerability. Despite the contract's reliance on extensive calculations to generate pseudo-random numbers, malicious miners can obtain the block information, including the timestamp, and disclose it to attackers. It may enable attackers to derive the random numbers from the calculation method in the source code and subsequently exhaust tokens from the contract.

\begin{figure}[ht]
  \centering
  \includegraphics[width=0.6\linewidth]{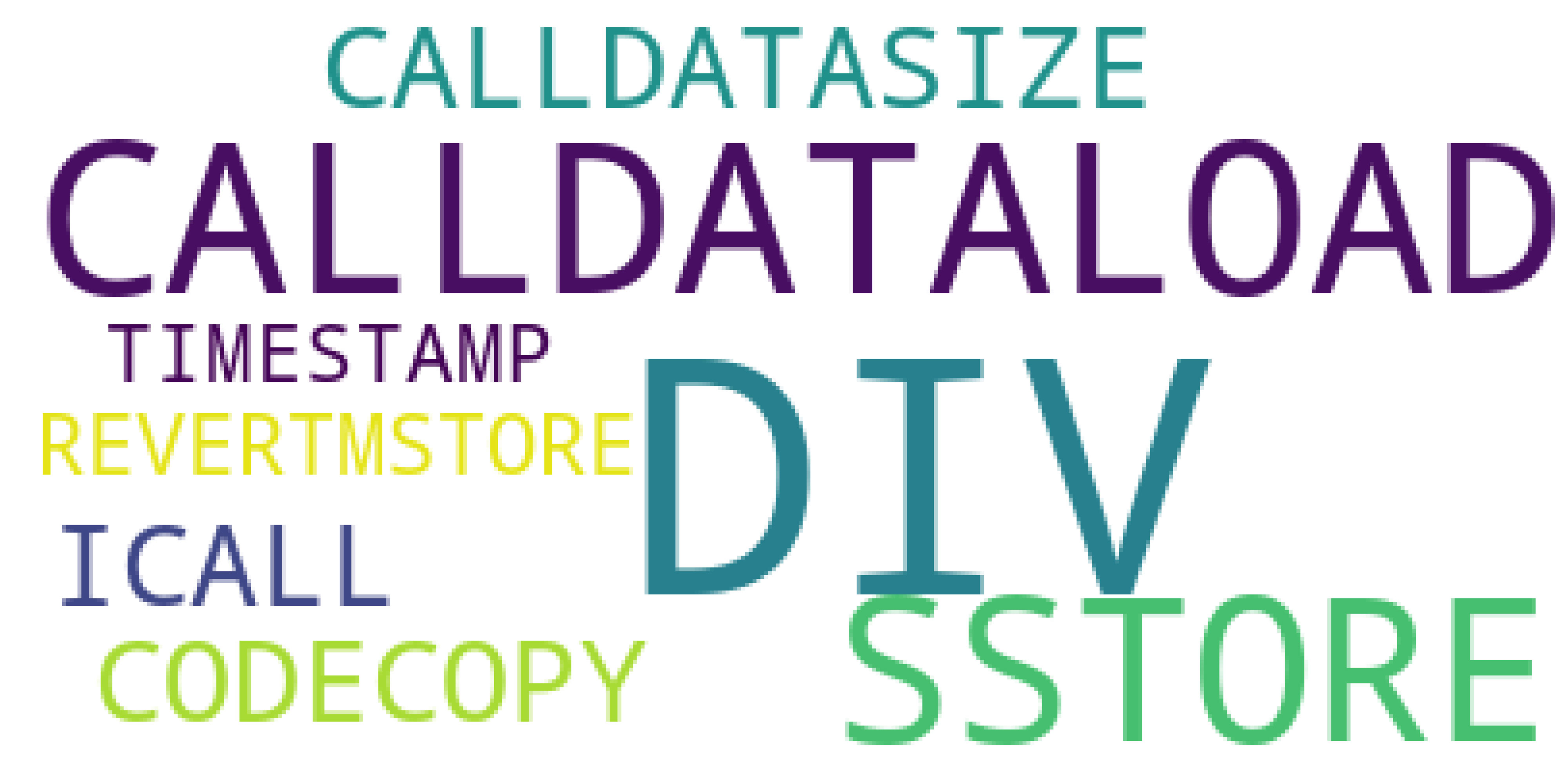}
  \caption{The Frequencies of SSA Opcodes in Time Manipulation Vulnerabilities}
  \label{fig:time_cloud_graph}
  \vspace{-1ex}
\end{figure}

In Figure \ref{fig:time_cloud_graph}, we present a summarization of the SSA Opcodes associated with the timestamp manipulation vulnerability. The Opcodes include \textsc{timestamp}, which indicates that the contract accesses the block dependency of the timestamp attribute; \textsc{calldataload}, which retrieves the call data; \textsc{sstore}, which stores data to storage; and \textsc{div}, which denotes data manipulation. These instructions indicate that contracts with these vulnerabilities are likely to read call data and timestamp, process the data, and modify the storage. By leveraging function properties, we can determine whether a function tends to modify state variables. Furthermore, we conducted a comprehensive analysis of all state variables. Our findings indicate that 90.42\% of contracts are not \textit{payable}. 92.57\% of the functions that contain the payable type modify the state variables.

\subsubsection{Unchecked Low-Level Call Vulnerability} Additionally, we have distinguished unchecked low-level call vulnerabilities in Listing \ref{lst:Integer}. Specifically, on line 16, the contract's transfer method utilizes the \texttt{send}, which is considered a low-level call. This approach sends tokens or ETHs to another contract address and returns a boolean value indicating the success or failure of the transaction. However, the contract does not include any checks on the transaction status, which will result in an incomplete or unsuccessful transaction. Although the vulnerability in this contract does not directly affect the balance of the contract, it can negatively impact users when the contract does not send tokens or ETHs. Therefore, we consider it to be a true positive case.

\begin{figure}[ht]
  \centering
  \includegraphics[width=0.6\linewidth]{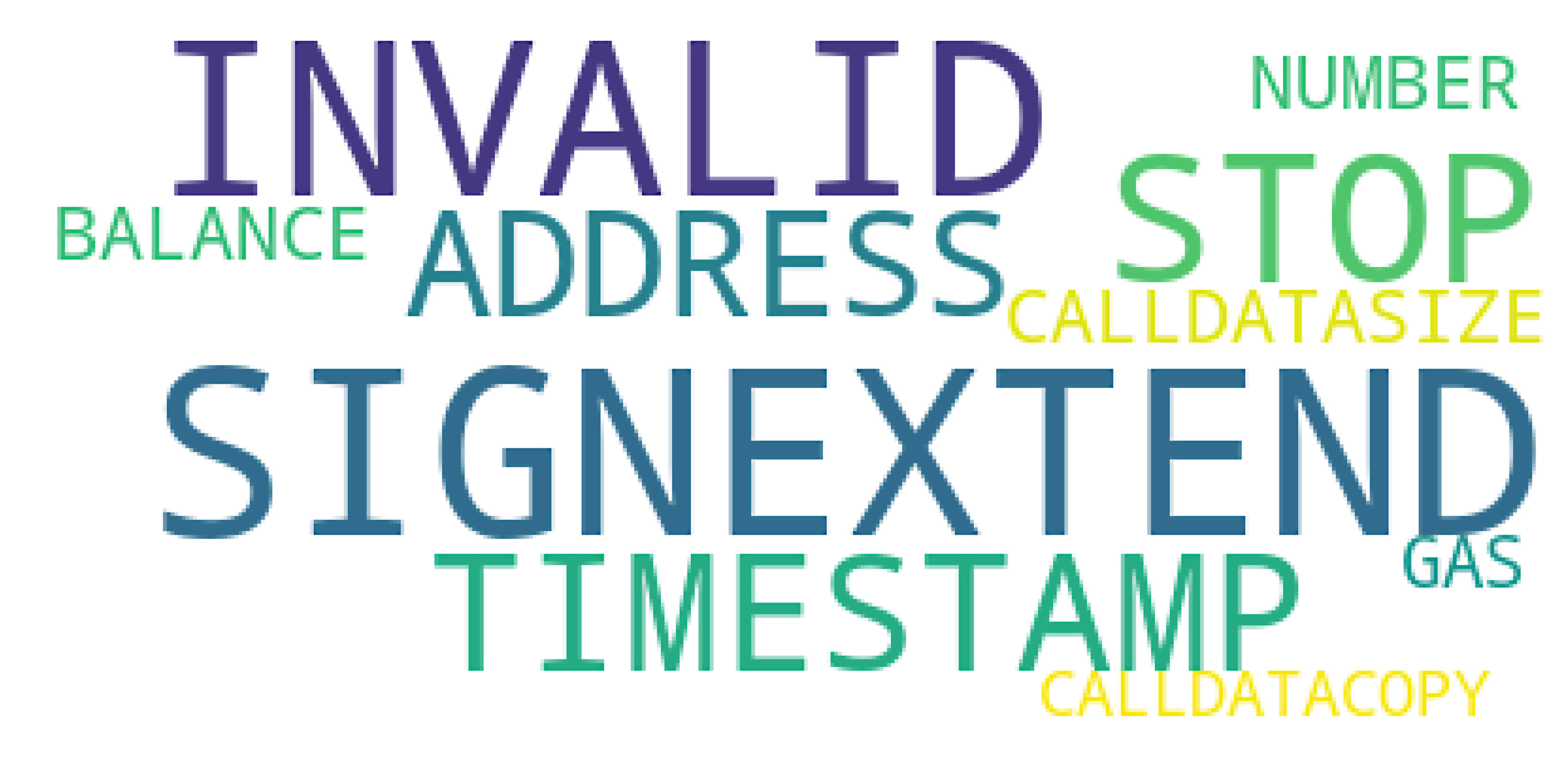}
  \caption{The Frequencies of SSA Opcodes in Unchecked-Low-Level Call Vulnerabilities}
  \label{fig:ullc_cloud_graph}
  \vspace{-2ex}
\end{figure}

We conducted an analysis of Solidity assembly Opcodes depicted in Figure \ref{fig:ullc_cloud_graph} and found that the \textsc{address} and \textsc{balance} Opcodes are frequently utilized in these contracts with unchecked-low-level calls. These Opcodes are instrumental in capturing the address and balance of contracts, respectively, and are often used in functions that read state variables. Furthermore, we surveyed a significant disparity of over 4.27 times between the number of payable and non-payable functions in the analyzed contracts. Among \textit{payable} functions, 54.68\% have modified the state, while 45.32\% have read state variables. 
% As a result, distinguishing vulnerabilities between \textit{view} and \textit{pure} function types poses a challenge, which will be discussed in the following.
At the Opcode level, low-level calls in Solidity can be categorized into three types: \textsc{call}, \textsc{delegatecall}, and \textsc{callcode}. However, in the semantic context gathered in the SSA format, \textsc{delegatecall} and \textsc{callcode} are not frequently detected. Consequently, analyzing fragile contracts containing such low-level calls (e.g., dangerous delegatecall) would yield true negatives.

\subsubsection{Transaction Ordering Dependency} Exploiting a transaction ordering dependency (TOD) vulnerability largely depends on the execution order. When the vulnerability occurs, indicating that the transaction is profitable, state variables and balance-related operations may be shared between the two transactions. Consequently, examining the \textsc{sstore} and \textsc{balance} Opcodes is necessary, as \textsc{sstore} stores the state variables, and \textsc{balance} retrieves the account's balance. Line 14 of Listing \ref{lst:Integer} also reveals the possibility of TOD. If transaction $t_1$ wins the bet by transferring the correct value, transaction $t_2$ could front-run the $t_1$. Mythril failed to identify the TOD, which caused the report to be a false positive.

\subsubsection{Reentrancy Vulnerability} 
\begin{figure}[ht]
  \centering
  \includegraphics[width=0.7\linewidth]{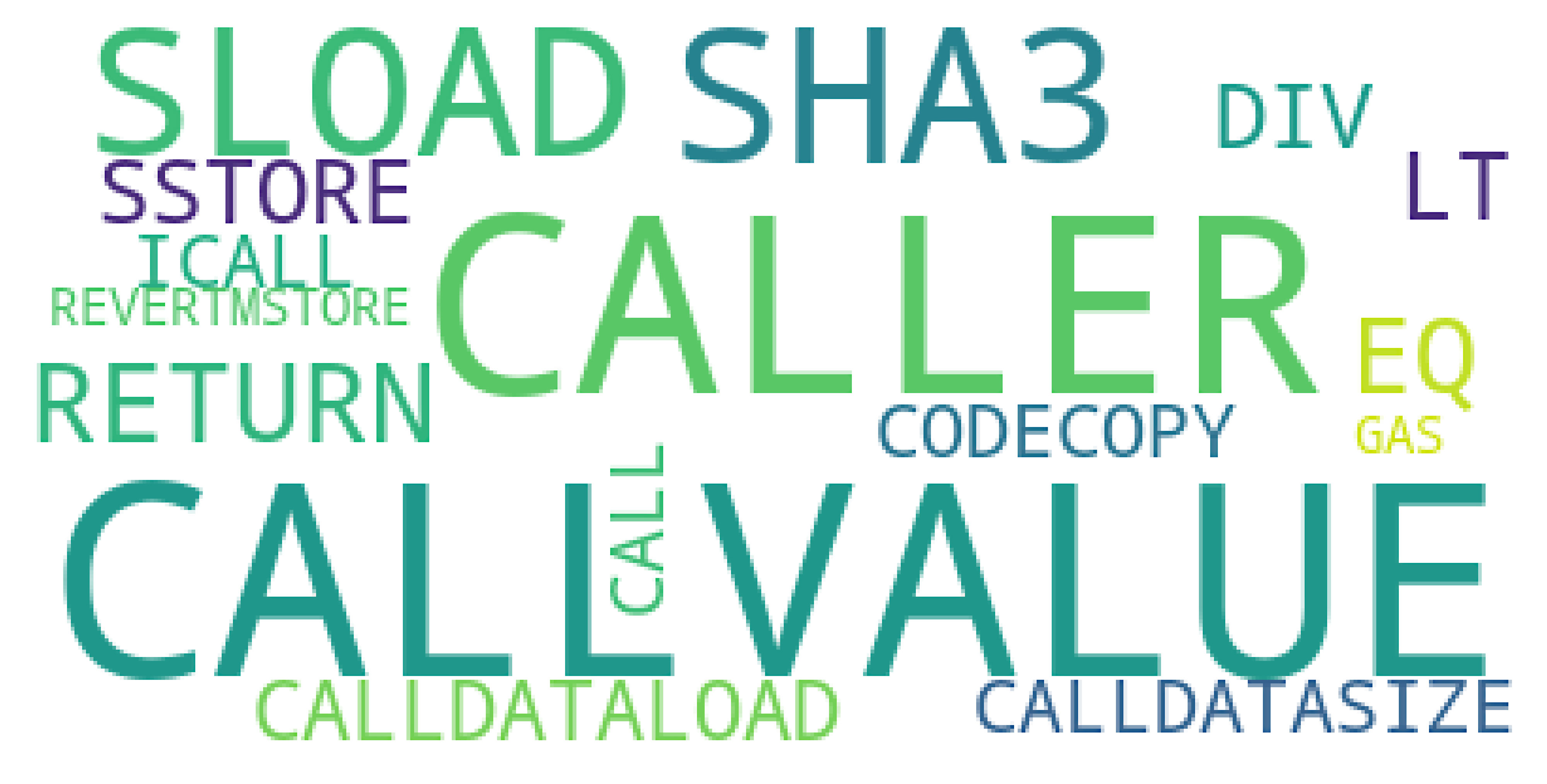}
  \caption{The Frequencies of SSA Opcodes in Reentrancy Vulnerabilities}
  \label{fig:reentrancy_cloud_graph}
\end{figure}

As described at \cref{subsec::rq4}, the reentrancy vulnerability in our database shows a tendency. In this issue, some operations, such as utilizing transfer and send methods, require the \textit{payable} type, which is associated with the \textsc{callvalue} instruction. Additionally, when using the \texttt{call.value}, the attacker can access the target function's signature by utilizing the \textsc{sha3} instruction. Furthermore, an examination of the reentrancy vulnerability reveals that certain Opcodes, such as \textsc{caller}, are frequently utilized. Consequently, as evidenced by the results depicted in Figure \ref{fig:reentrancy_cloud_graph}, it is reasonable to conclude that \texttt{msg.caller} and \texttt{msg.value} are commonly present in the vulnerable contract.

An illustration of a potential contract vulnerability is presented in Listing \ref{lst:Reentrancy}, which could be exploited through reentrancy. Specifically, when tokens are sent to an address without checking the balance, attackers can repeatedly call the function and exhaust its resources. To address this issue, protective measures, such as utilizing transfer and send methods, require the \textit{payable} type, which is associated with the \textsc{callvalue} instruction. Additionally, when using the \texttt{call.value} function call, the attacker can access the target function's signature by utilizing the \textsc{sha3} instruction. Furthermore, an examination of the reentrancy vulnerability reveals that certain Opcodes, such as \textsc{caller}, are frequently utilized. Consequently, as evidenced by the results depicted in Figure \ref{fig:reentrancy_cloud_graph}, it is reasonable to conclude that \texttt{msg.caller} and \texttt{msg.value} are commonly present in the vulnerable contract. 

\begin{lstlisting}[language=Solidity,
                    basicstyle=\footnotesize\tt,
                    numbers=left,
                    captionpos=b,
                    belowskip = -1em,
                    caption= The Simplified Snippets of \textbf{PrivateBank},
                    label=lst:Reentrancy,
                    ] 
  contract PrivateBank {
    mapping (address=>uint) public balances;
    function CashOut(uint _am) {
      if(_am<=balances[msg.sender]) {            
        if(msg.sender.call.value(_am)()){
          balances[msg.sender] -= _am; }
      }
    }
  }
\end{lstlisting}

\subsection{Limitations}\label{subsec::lim}
This section describes the limitations of our research, which is restricted by time, reverse engineering, and raw data.

\subsubsection{Time} The two-stage framework is presented for detecting vulnerabilities in bytecode-level smart contracts, which is necessary due to the limited availability of ABI information. In the presence of raw ABI data, we can extract the ABIs from Etherscan, while in the absence of raw ABI data, we employ the \textsc{SRIF} function to infer the function signatures and properties of the contracts. However, this inferred data does not contain the output parameters and types of functions. The extraction of ABI, function parameters, and attributes must be completed before the execution of \textsc{COBRA}, with function parameters and attributes being inferred using the \textsc{SRIF}. As a result, the construction of the framework requires a great deal of time. We have taken the first step toward resolving this issue by proposing a deep learning-based framework for function signature inference. In the future, we will combine this method with our encoder to build an end-to-end deep learning framework for improved vulnerability detection.

 % In the experiments described in this paper, we only employed 99,745 function signatures that were not fine-tuned because our hardware and time resources were primarily devoted to identifying vulnerabilities. In addition, the proposed framework is two-stage because it is not possible to obtain the ABI of all contract simultaneously.

\subsubsection{Reverse Engineering} For the structure of \textsc{SRIF} and \textsc{COBRA}, they both rely on reverse engineering of the EVM. First, they might be affected by the accuracy of Opcode acquisition. For Opcode, we use the \textsc{pyevmasm} module to facilitate integration development directly in Python programming language. However, this module also results in a loss of precision. It does not support all versions of \textsc{solc}, which is the Solidity compiler in Ethereum. On the other hand, \textsc{SRIF} also relies on CFG recovery in both original and SSA format Opcodes. In our implementation, the CFG module in EtherSolve is utilized for inferring a more precise CFG, which verifies whether the CFG contains unreachable basic blocks to improve accuracy. Although we could improve the accuracy of CFG and reduce the impact of noise to a certain extent, the volume of data required for deep learning in \textsc{SRIF} and \textsc{COBRA} is still substantial.

\subsubsection{Raw Data} In the absence of raw ABI, our approach is limited to providing alerting functionality only. In the 4byte signature library, there are far more than 1 million unique records. Nevertheless, according to our survey, only around 15.62\% of 96,200 contracts are publicly available to ABI. As the experiments (\cref{sec:experiments}) demonstrated, even though our function parameter inference method \textsc{SRIF} could achieve relatively good results, there is still an apparent limitation in vulnerability detection compared to the raw ABI. Therefore, we have a relatively reduced capacity for detecting contracts that keep ABI private. However, as the recovery capability of the function interface is optimized and the amount of ABI disclosed increases, we would have a significant potential for improving our approach.
% Another significant limitation is the lack of precisely labeled vulnerability datasets in bytecode smart contracts. In the SWC and smartbugs databases, only the source code data of the relevant vulnerabilities are publicly available, and no bytecode data is available. Therefore, we cannot guarantee that these data will be valid on the ground truth data. Since our labeled data is from smartbugs, we can just guarantee that our detection results will be close to smartbugs.

\subsection{Threat to Validity}
\label{subsec:validity}
\noindent\textbf{Internal validity.} The contracts in XBlock\_ETH are from the Ethereum blockchain, where the ground truth labels for function signatures are derived from 4bytes (See \cref{sec:subsub_datasets}), collected from real-world functions. Therefore, it is suitable for evaluation. 
Additionally, COBRA relies on reverse engineering the Ethereum virtual machine. For bytecode data, it needs to be converted to an opcode sequence. However, there are errors in the current method of obtaining opcode sequences, and different versions of compilers will lead to accuracy loss. Even some popular compiler tools, such as pyevmasm~\cite{crytic2020}, do not support all versions, resulting in a lack of accuracy. However, COBRA exploits an intermediate representation in the SSA form and thus alleviates inaccurate translations of opcode sequences. In addition, the EtherSolve tool \cite{contro2021ethersolve} is used in this paper to achieve the accurate construction of the CFG graph to reduce invalid basic blocks in the recovery process.

\noindent\textbf{External validity.} SRIF has some randomness when compared with Gigahorse. We split the test set into equal numbers of groups and randomly selected contracts from each group, ensuring as fair a comparison as possible.

\vspace{-1ex}
\subsection{Future Direction}
This section outlines potential enhancements for subsequent research. First, our current implementation of SRIF focuses solely on input parameters while excluding return value specifications in function signatures. The ablation study presented in §IV-E reveals that incorporating ABI output data enhances vulnerability detection recall by approximately 2.06 percentage points. This finding motivates planned extensions to integrate output parameter analysis into SRIF's framework. Second, while both SRIF and COBRA employ recurrent neural network architectures for computational efficiency rather than the Transformer-based models, the emergence of advanced large language models (LLMs) presents new opportunities. Future investigations will evaluate the feasibility of adapting state-of-the-art LLMs (e.g., GPT-4, Gemini, Claude) for simultaneous function signature inference and vulnerability identification, contingent upon overcoming current problems, including computational constraints, model robustness, error bounds, and explainability of predictions.

\section{Related Work}
\label{sec:related_work}
% In this section, we describe related works on vulnerability detection with machine learning and function signature recovery.

\subsection{Smart Contract Vulnerability Detection}
Many state-of-the-art works based on machine learning have been presented for vulnerability detection \cite{chen2018detecting, he2019ILFuzzer, gao2020deep, So2021smarTest, huang2021hunting, sendner2023smarter,li2024detecting, kong2025uechecker,zou2025malicious,li2021hybrid,niu2025natlm, kong2023defitainter, liu2022finding, ghaleb2023achecker, chen2024improving, wang2024contractcheck,  chen2025smarttrans, liu2025anomaly, wang2025tmf, xie2025block, huang2025deep}, as the increase in smart contract bugs and machine learning technology. 
Chen et al. \cite{chen2018detecting} use machine learning methods to detect Ponzi schemes. Ether flow graphs are constructed by analyzing Opcode and account information, and features are designed for classifying source code contracts. He et al. \cite{he2019ILFuzzer} implement the fuzz function through the GRU module and combine it with symbolic execution techniques to achieve higher coverage. Gao et al. \cite{gao2020deep} transform the code into an abstract syntax tree (AST) and then serialize the tree based on the nodes. After learning the feature vector of the sequence, the vector threshold is used to determine whether the feature is a vulnerability. So et al. \cite{So2021smarTest} collect enough sequences of vulnerabilities through symbolic execution to train a language model. The tool detects Ether leaking and suicidal contracts in source code. Huang et al. \cite{huang2021hunting} utilized an unsupervised graph embedding algorithm to embed the sliced CFG in the graph and then performed similarity calculations on the sliced vectors. Sendner et al. \cite{sendner2023smarter} proposed ESCORT, a multi-label detection tool that supports lightweight transfer learning. Different from these above works, COBRA takes the advantages of the function interface into account when detecting vulnerabilities in smart contracts.

\vspace{-1ex}
\subsection{Function Signature Recovery}
Different from the Java Virtual Machine (JVM), EVM does not retain function signature information in bytecode data. Call data stores function parameters that can only be accessed via a particular processing. Numerous ways utilize these databases to recover function signatures by developing parameter acquisition procedures (e.g., Gigahorse \cite{grech2019gigahorse}, Eveem \cite{kolinko2018eveem}). Gigahorse introduces the "\textsc{callprivate}" directive to identify private function calls. Eveem, utilizing symbolic execution techniques, performs symbolic and algebraic computations of the execution trace. When a layout containing \textit{offset} and \textit{num} fields is found, it is regarded as an array. SigRec \cite{chen2021sigrec} searches for call data in execution traces related to \textsc{calldatacopy} and \textsc{calldataload} and creates specific inference rules to recover various function signatures.
Furthermore, there are also methods for locating function information without establishing rules (e.g., OSD \cite{ethervm2023OSD}, Neural-FEBI \cite{he2023neural}, and DeepInfer \cite{zhao2023deepinfer}). Typically, OSD searches directly in EFSD \cite{4byte} for function hashes to discover function signatures. Neural-FEBI identifies functions via a two-step process to get a more precise CFG.

% \vspace{-2ex}
\section{Conclusion}
\label{sec:conclusion}

% We propose \textsc{COBRA} with a novel encoder and the \textsc{SRIF}. \textsc{COBRA} combines semantic and function features of contracts and generates an implicit feature expression for detecting bytecode smart contract vulnerabilities in Ethereum. In addition, the \textsc{SRIF} was first proposed as function signature recovery component to address the case where ABI information is not disclosed. Specifically, we make a minor adjustment to MS-CAM so that it can learn both global and local features for each function. Experiments demonstrate that the function signature recoverer can accurately and efficiently predict function parameters for a 99,745 signatures dataset. Encoders equipped with publicly accessible ABI have efficient vulnerability detection capabilities. As a reminder without publicly available ABI, the model can achieve a recall of over 90\%. Therefore, the recovery of ABI information must continue to be addressed, and more effective methods should be examined in the future.

We present \textsc{COBRA}, a novel framework for detecting vulnerabilities in Ethereum smart contracts at the bytecode level. COBRA employs semantic and function interface features for vulnerability detection. Additionally, we introduce the \textsc{SRIF}, a function signature recovery technique that handles cases where ABI is not disclosed. We also conduct minor adjustments to \textsc{ms-cam} to learn both global and local features for functions in each contract. Our experiments demonstrate that the SRIF can accurately and efficiently predict function parameters, achieving a 94.76\% F1-score on the dataset of 99,745 signatures. \textsc{COBRA} equipped with publicly available ABI exhibit 93.45\% F1-score in vulnerability detection. However, even without publicly available ABI, the recall rate remains above 89\%. Therefore, recovering ABI information remains a crucial consideration, and further research into more effective techniques is necessary.

% \section{ACKNOWLEDGMENTS}
% This work is sponsored by the National Natural Science Foundation of China (No.62362021 and No.62402146), CCF-Tencent Rhino-Bird Open Research Fund (No.RAGR20230115), and Hainan Provincial Department of Education Project (No.HNJG2023-10).

\normalem
\bibliographystyle{IEEEtran}

\bibliography{main}

% \balance
% with photo
\vspace{-2em}
\begin{IEEEbiography}[{\includegraphics[width=1in,height=1.25in,clip,keepaspectratio]{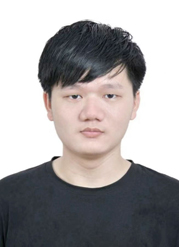}}]{Wenkai Li} is currently pursuing the doctor’s degree in the School of Cyberspace Security at Hainan University, China. Previously, he received a master's degree in the School of Cyberspace Security at Hainan University. His research lies in smart contract security and malicious behavior analysis, focusing on enhancing blockchain security through software and data analytics. He is also exploring the integration of artificial intelligence, such as graph neural networks and large language models.
\end{IEEEbiography}
\vspace{-2em}

\begin{IEEEbiography}[{\includegraphics[width=1in,height=1.25in,clip,keepaspectratio]{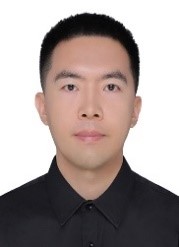}}]{Xiaoqi Li}
is an associate professor at Hainan University. Previously, he was a researcher at the Hong Kong Polytechnic University. He received his Ph.D. in Computer Science from Hong Kong Polytechnic University, MSc in Information Security from the Chinese Academy of Sciences, and BSc in Information Security from Central South University. His current research interests include Blockchain/Mobile/System Security and Privacy, Ethereum/Smart Contract, Software Engineering, and Static/Dynamic Program Analysis. He received best paper awards from INFOCOM’18, ISPEC’17, CCF’18, and an outstanding reviewer award from FGCS’17.
\end{IEEEbiography}
\vspace{-2em}
\begin{IEEEbiography}[{\includegraphics[width=1in,height=1.25in,clip,keepaspectratio]{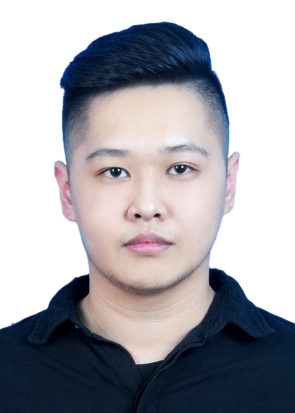}}]{Yingjie Mao}
is currently pursuing a master’s degree in the School of Cyberspace Security at Hainan University, China. Previously, he received a B.E. degree from the Southwest University of Science and Technology. His current research interests include Blockchain Security/Privacy and Large Language Model.
\end{IEEEbiography}
\vspace{-2em}
\begin{IEEEbiography}[{\includegraphics[width=1in,height=1.25in,clip,keepaspectratio]{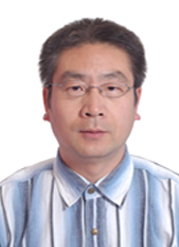}}]{Yuqing Zhang}
is the Director of the Chinese National Computer Network Intrusion Prevention Center, Deputy Director of the Chinese National Engineering Laboratory of Computer Virus Prevention Technology, Vice Dean of the School of Computer and Control Engineering at the Chinese Academy of Sciences, and Professor at Hainan University. He received his Ph.D. from Xi'an University of Electronic Science and Technology. He has presented over 100 papers and 7 national/industry standards. His current research interests include Network Attacks and Prevention, Security Vulnerability Mining and Exploitation, IoT System Security, AI Security, Data Security, and Privacy Protection. 
\end{IEEEbiography}

\vfill
\end{document}